\documentclass[prb,twocolumn,showpacs,preprintnumbers]{revtex4-1}
\usepackage{latexsym,epsfig,amssymb,amsfonts,amsmath,graphicx,color,bm,times,amstext,epstopdf,braket,float,ulem,bbm}

\def\a{\alpha}

\def\d{\delta}
\def\e{\epsilon}

\def\h{\eta}

\def\S{\Sigma}
\def\s{\sigma}

\def\w{\omega}
\def\ua{\uparrow}
\def\da{\downarrow}

\def\Vec#1{\mathbf #1}

\newcommand{\Snor}{\Sigma^\text{nor}}
\newcommand{\Sano}{\Sigma^\text{ano}}
\newcommand{\Akw}{A(\Vec{k},\omega)}
\newcommand{\kAN}{\Vec{k}_\text{AN}}
\newcommand{\Tc}{T_\text{c}}
\newcommand{\ef}{\epsilon_f}
\newcommand{\wmott}{\w_\text{Mott}}
\newcommand{\wwfa}{\w_\text{WF1}}
\newcommand{\wwfb}{\w_\text{WF2}}
\newcommand{\fmott}{f^\text{Mott}}
\newcommand{\fwf}{f^\text{WF}}

\begin{document}

\title{Direct Connection between Mott Insulator and $d$-Wave High-Temperature Superconductor Revealed by Continuous Evolution of Self-Energy Poles}

\author{Shiro Sakai$^1$, Marcello Civelli$^2$, and Masatoshi Imada$^3$}

\affiliation{
$^1$Center for Emergent Matter Science, RIKEN, Wako, Saitama 351-0198, Japan\\
$^2$Laboratoire de Physique des Solides, CNRS, Univ. Paris-Sud, Universit\'e Paris-Saclay, 91405 Orsay Cedex, France\\
$^3$Department of Applied Physics, University of Tokyo, Hongo, Tokyo 113-8656, Japan
}
\date{\today}
\begin{abstract}
The high-temperature superconductivity in copper oxides emerges when carriers are doped into the parent Mott insulator. This well-established fact has, however, eluded a microscopic explanation. Here we show that the missing link is the self-energy pole in the energy-momentum space. Its continuous evolution with doping directly connects the Mott insulator and high-temperature superconductivity. We show this by numerically studying the extremely small doping region close to the Mott insulating phase in a standard model for cuprates, the two-dimensional Hubbard model. 
We first identify two relevant self-energy structures in the Mott insulator; the pole generating the Mott gap and a relatively broad peak generating the so-called waterfall structure, which is another consequence of strong correlations present in the Mott insulator.
We next reveal that either the Mott-gap pole or the waterfall structure (the feature at the energy closer to the Fermi level) directly transforms itself into another self-energy pole at the same energy and momentum when the system is doped with carriers. The anomalous self-energy yielding the superconductivity is simultaneously born exactly at this energy-momentum point.
Thus created self-energy pole, interpreted as arising from a hidden fermionic excitation, continuously evolves upon further doping and considerably enhances the superconductivity.
 Above the critical temperature, the same self-energy pole generates a pseudogap in the normal state. We thus elucidate a unified Mott-physics mechanism, where the self-energy structure inherent to the Mott insulator directly gives birth to both the high-temperature superconductivity and pseudogap.   
\end{abstract}
\maketitle

\section{Introduction}

High-temperature superconductivity in copper oxides occurs when carriers are doped into a parent Mott insulator \cite{bednorz86}.
This observation \cite{anderson87} has brought about enormous number of studies on the role of strong electronic correlations in the high-temperature superconductivity.
In fact, various theoretical studies, including numerical simulations which seriously take into account the strong correlation effect, have shown that superconductivity is a strong candidate of low-temperature phases in the doped Mott insulators\cite{maier05RMP,gull13PRL,misawa14}.
Nevertheless, how the Mott insulator transforms into the superconductor by doping ($\d$), and why a high transition temperature ($\Tc$) results from the Mott physics are questions that remain still open.

Another unresolved issue of the cuprates is the anomalous metallic state observed above $\Tc$. Especially in a lightly-doped region, a gap called ``pseudogap'' has been observed in various experiments \cite[and references therein]{alloul89,warren89,yasuoka89,timusk99,fujita12,hashimoto14}. Its relation to the superconductivity as well as to the Mott insulator, has also been controversial despite an intensive debate in the last few decades.

All these three states, i.e., the Mott insulator, pseudogap metal, and high-$\Tc$ superconductor, constitute nontrivial electronic states which defy the description by the standard theories of solids \cite{imada98}, like the band theory, Fermi-liquid theory, and Bardeen-Cooper-Schrieffer theory \cite{bardeen57}. 
One of the simplest models that have been proposed to accommodate all these three states is the two-dimensional Hubbard model \cite{anderson87}, which well takes into account the electronic correlations resulting from the local Coulomb interaction. 
Recent numerical studies, based on quantum cluster theories \cite{hettler98,lichtenstein00,senechal00,kotliar01,potthoff03,maier05RMP}, have
revealed a presence of self-energy poles in these three states\cite{maier02,senechal04,civelli05,stanescu06,berthod06,kyung06PRB1,haule07,maier08,kyung09,sakai09PRL,civelli09PRL,liebsch09,ferrero09,sakai10,gull15,sakai16,sakai16PRB}. This fact well explains the inapplicability of the above-mentioned standard theories, which cannot describe the singular self-energy.

\begin{figure}[!htb]
\center{
\includegraphics[width=0.48\textwidth]{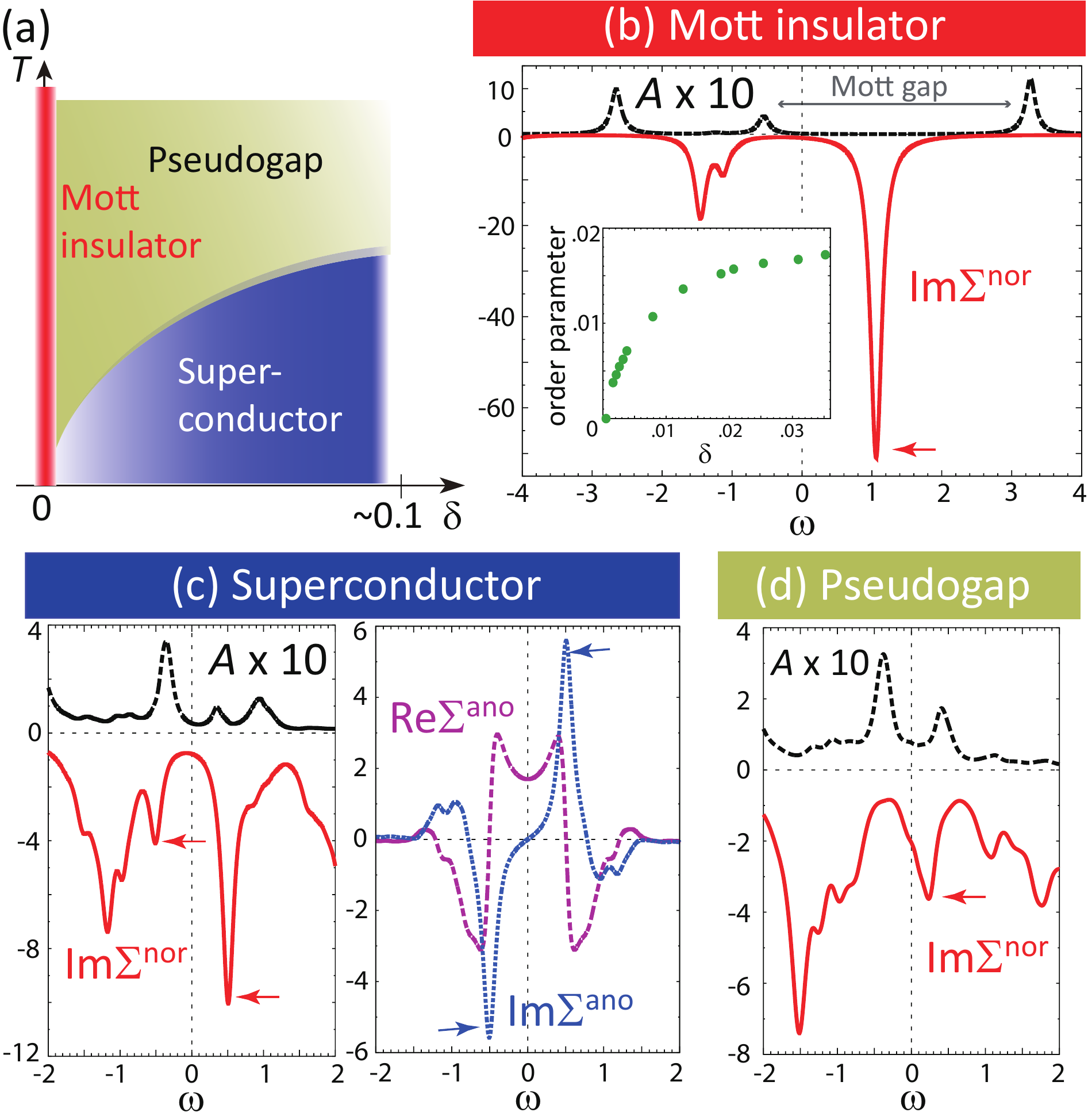}}
\caption{(a) Schematic phase diagram of the two-dimensional Hubbard model with a strong repulsive interaction in the vicinity of half filling. 
(b) $A(\kAN,\w)$ and Im$\Snor(\kAN,\w)$ with $\kAN\equiv(\pi,0)$ in the Mott-insulating state ($\d=0, T=0.01$), where the Fermi level ($\w=0$) is taken to be just above the occupied states.  $t=1$ and $U=8$ are used for (b)-(d).
Inset plots the superconducting order parameter against $\d$, calculated at $T=0.01$.
(c) $A$, Im$\Snor$ (left) and $\Sano$ (right) in the superconducting state ($\d=0.048, T=0.01$).
(d) $A$ and Im$\Snor$ in the pseudogap state ($\d=0.047, T=0.06$).
The arrows denote the self-energy peaks discussed in the text.}
\label{fig:phase}
\end{figure}

In this paper, we reveal the microscopic relation between these Mott insulator, high-$\Tc$ superconductivity and pseudogap, especially between the first two phases, by exploring how the self-energy evolves with doping the Mott insulator.
In the superconducting state, the self-energy pole appears also in its anomalous component, and it ultimately enhances the superconducting pairing \cite{maier08}.
We find that this self-energy pole enhancing the superconductivity has the root at the self-energy peaks present in the Mott insulator. 
This result validates and substantiates the long-standing but still-speculative argument that the Mott-insulating state at zero doping is at the origin of the high-$\Tc$ superconductivity, revealing the microscopic mechanism in terms of the self-energy structure.

Figure \ref{fig:phase}(a) schematically illustrates the doping-temperature ($T$) phase diagram of the two-dimensional Hubbard model, obtained by quantum-cluster theories \cite{hettler98,lichtenstein00,senechal00,kotliar01,potthoff03,maier05RMP}, close to half filling ($\d=0$) and at intermediate-to-strong coupling\cite{maier05RMP,kancharla08,sordi12PRL,gull13PRL,sakai16PRB,fratino16,reymbaut16}.
The Mott-insulating state appears at $\d=0$, where the self-energy shows a prominent pole [at $\w\simeq 1$ in Fig.~\ref{fig:phase}(b)].
This pole generates the Mott gap (for $0\lesssim\w\lesssim3$) between the occupied and unoccupied weights in the spectral function $A(\Vec{k},\w)$. 

Carrier doping immediately alters the Mott insulator into a superconductor at low temperatures, as displayed in the inset to Fig. 1(b) by a non-zero 
value of the superconducting order parameter, $\langle c_{i\ua} c_{j\da}\rangle$ with $i$ and $j$ denoting the nearest neighboring sites, which respects the $d_{x^2-y^2}$ symmetry of the pairing.
This immediate emergence of superconductivity is not the case of the cuprates, where it appears only after a substantial hole doping ($\gtrsim 5\%$). 
The latter is likely due to the inhomogeneity inherent to the real materials and a remnant antiferromagnetic order.
In this paper, we assume the translational invariance and charge uniformity of the system, to study a continuous doping evolution from the Mott insulator to superconductor in a clean system.

In this superconductor, both the normal ($\Snor$) and anomalous ($\Sano$) components of the self-energy show poles at low energies [Fig.~\ref{fig:phase}(c)] \cite{poiblanc02,poiblanc03,poiblanc05,haule07,maier08,kyung09,civelli09PRL,gull15,sakai16}.
As their energy positions perfectly agree, they share the same origin \cite{gull15,sakai16}.
A Kramers-Kr\"onig analysis shows that this pole considerably lifts the low-energy value of Re$\Sano$, enhancing the superconductivity \cite{poiblanc05,maier08,kyung09,civelli09PRL,sakai16,reymbaut16}. 

For $T>\Tc$ [Fig.~\ref{fig:phase}(d)], $\Sano$ vanishes while the low-energy peak of Im$\Snor$ persists, yielding a small gap in $A(\Vec{k},\w)$.
This gap has been identified with the pseudogap observed in the cuprates above $\Tc$\cite{maier02,stanescu06,kyung06PRB1,sakai09PRL,liebsch09,sakai10,eder11,wu17,braganca17}.
Though this peak of Im$\Snor$ is broadened by thermal fluctuations, it evolves into a pole at low $T$ in the superconducting phase [as displayed in Fig.~\ref{fig:phase}(c)] \cite{sakai16} and also in the normal paramagnetic state if this latter is imposed at low $T$ by constraining the CDMFT equations \cite{dzyaloshinskii03,kyung06PRB1,konik06,berthod06,yang06,stanescu06,stanescu07,liebsch09,sakai09PRL,sakai10,eder11,wu17,braganca17}.

In Ref.~\onlinecite{sakai16}, we showed the continuity of this self-energy peak through $\Tc$, which implies the same origin of the pseudogap and high-$\Tc$ superconductivity.
Furthermore, we found that, in the single-particle Green's function, the pole of $\Snor$ cancels with the contribution from the anomalous part. We then revealed that this cancellation means a presence of a hidden {\it fermionic} excitation coupling to the electron, and that the coupling generates the self-energy pole. The hidden fermion emerges from a strong correlation effect while its explicit identity is yet to be clarified.


\section{Model and method}
We consider the two-dimensional Hubbard model,
\begin{align}
H= \sum_{\Vec{k}\s}\e(\Vec{k})c_{\Vec{k}\s}^\dagger c_{\Vec{k}\s}^{\phantom {\dagger}}+ U\sum_{i}n_{i\ua}n_{i\da},
\label{eq:hubbard}
\end{align}
where $c^{\phantom \dagger}_{\Vec{k}\s}$ ($c_{\Vec{k}\s}^\dagger$) annihilates (creates) an electron of spin $\s$ with momentum $\Vec{k}$ and $n_{i\s}$ is the density operator at site $i$ on a square lattice.
$U$ is the onsite Coulomb repulsion, and
$\e(\Vec{k})=-2t(\cos k_x+\cos k_y)-4t' \cos k_x \cos k_y -\mu$ with $t$ ($t'$) denoting the (next-)nearest-neighbor transfer integral and $\mu$ denoting the chemical potential.
We use $t=1$ as the unit of energy and $t'=-0.2$ throughout the paper. In real cuprates, the value of $t$ is estimated to be $\sim0.4-0.5$~eV by {\it ab initio} calculations\cite{hirayama17}.

Within the CDMFT \cite{kotliar01}, we map the model (\ref{eq:hubbard}) onto an effective impurity model consisting of a 2$\times$2 interacting-site cluster and eight noninteracting bath sites \cite{kancharla08,kyung09,civelli09PRB}.
We solve the latter model with a finite-$T$ extension\cite{liebsch05,capone07,liebsch12,sakai15} of the exact diagonalization method \cite{caffarel94}. 
This method allows us to calculate precise real-frequency properties, resolving changes with tiny dopings.
Unless otherwise mentioned, we set $T=0.01$ at which only the ground state of the effective impurity problem has a substantial Boltzmann weight.
The CDMFT outputs $\Snor$ and $\Sano$, which are related to the retarded Green function $\hat{G}$ in the Nambu-matrix form as  
\begin{align}
  \hat{\Sigma}(\Vec{k},\w)&=
 \left(
   \begin{array}{cc}
    \Snor(\Vec{k},\w) & \Sano(\Vec{k},\w)\\
     \Sano(\Vec{k},\w) & \Snor(\Vec{k},-\w)^\ast
    \end{array}
   \right) \nonumber \\
   &=
    \left(
    \begin{array}{cc}
     \w-\e(\Vec{k}) & 0\\
     0 & \w+\e(\Vec{k})
    \end{array}
   \right) - \left[\hat{G}(\Vec{k},\w)\right]^{-1}.
\label{sig}
\end{align}
Here we consider a spin-singlet superconductivity, for which we can choose
the phase of the two offdiagonal components to be the same. 

The normal part of the single-particle Green's function is given by the diagonal component of $\hat{G}$, i.e., 
\begin{align}
[\hat{G}(\Vec{k},\w)]_{11}=[\w-\e(\Vec{k})-\S^\text{nor}(\Vec{k},\w)-W(\Vec{k},\w)]^{-1}
\label{eq:G},
\end{align}
with 
\begin{align}
W(\Vec{k},\w)=\frac{\S^\text{ano}(\Vec{k},\w)^2}{\w+\e(\Vec{k})+\S^\text{nor}(\Vec{k},-\w)^*}.
\label{eq:W}
\end{align}
With these equations, the cancellation by the hidden fermion, mentioned in the last paragraph of the previous section, is explicitly expressed as a cancellation of the poles of $\Snor$ and $W$.\cite{sakai16}

The spectral function is defined by $\Akw \equiv -\frac{1}{\pi}\text{Im}[\hat{G}(\Vec{k},\w)]_{11}$. 
Its integral over momentum gives the density of states.
Because the pseudogap and the superconducting gap are most prominent in the antinodal region, we mainly study quantities at $\Vec{k}=\kAN\equiv (\pi,0)$, which can be derived directly from the cluster-impurity model, as explained in Ref.~\onlinecite{sakai16PRB}. We have used a $G$-periodization scheme \cite{kyung06PRB1} when we show the momentum-interpolated self-energy.
In order to plot the real-frequency properties, we use an energy-broadening factor $i\h$ with $\h=0.1$. 
The doping concentration $\d$ is calculated with the exact diagonalization method for the effective impurity model.
For more details of the method, we refer the readers to Ref.~\onlinecite{sakai16PRB}.

\section{Results}\label{sec:result}
\subsection{Electronic structure of Mott insulator}
\begin{figure}[tb]
\center{
\includegraphics[width=0.48\textwidth]{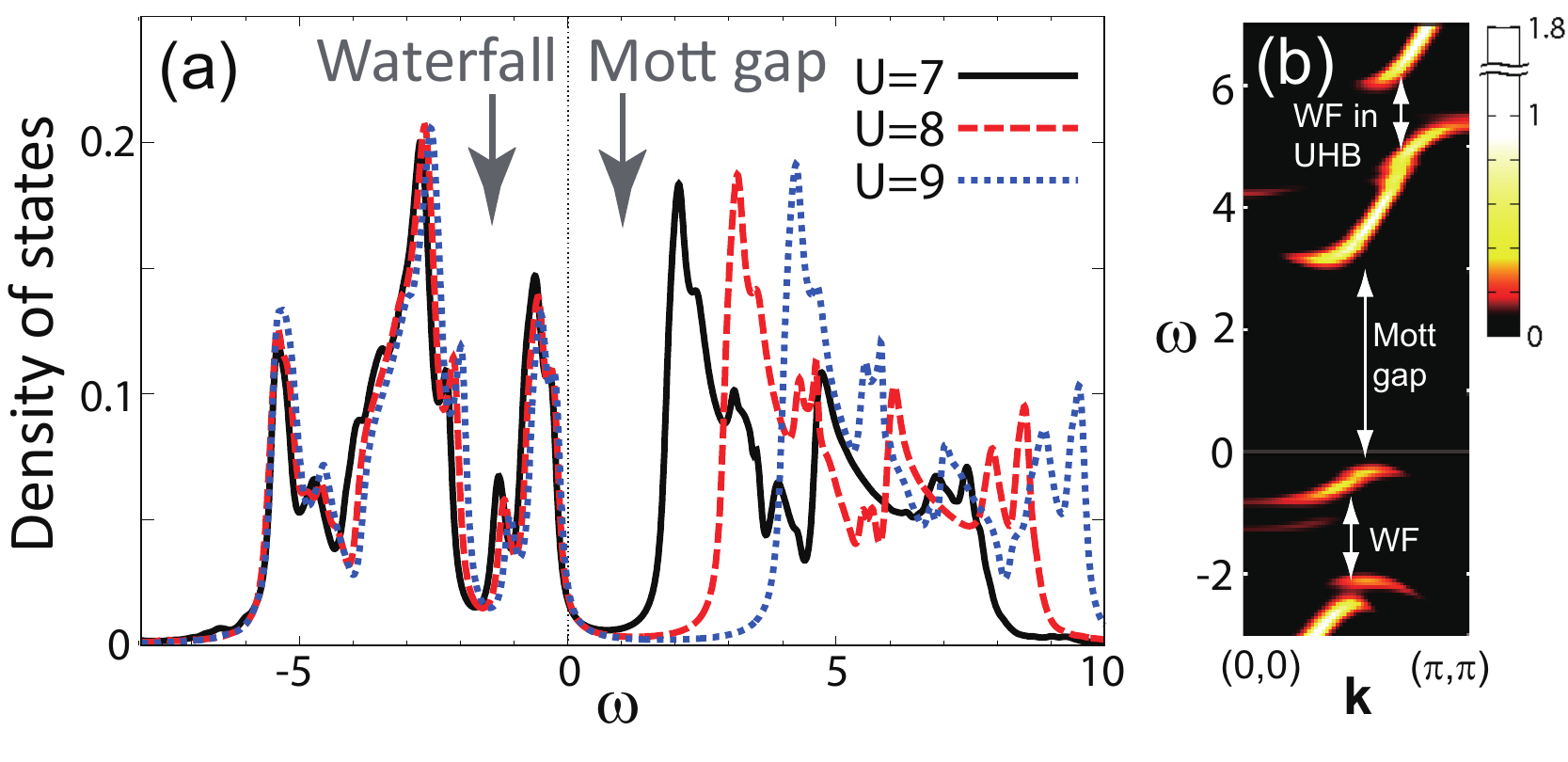}}
\caption{Electronic structure of the Mott insulator at $\d=0$ and $T=0.01$. (a) The density of states for $U=7, 8$ and $9$. The Fermi level is taken to be just above the occupied states. (b) The intensity plot of the spectral function for $U=8$ along the $(0,0)-(\pi,\pi)$ line.}
\label{fig:dos}
\end{figure}

\begin{figure}[tb]
\center{
\includegraphics[width=0.48\textwidth]{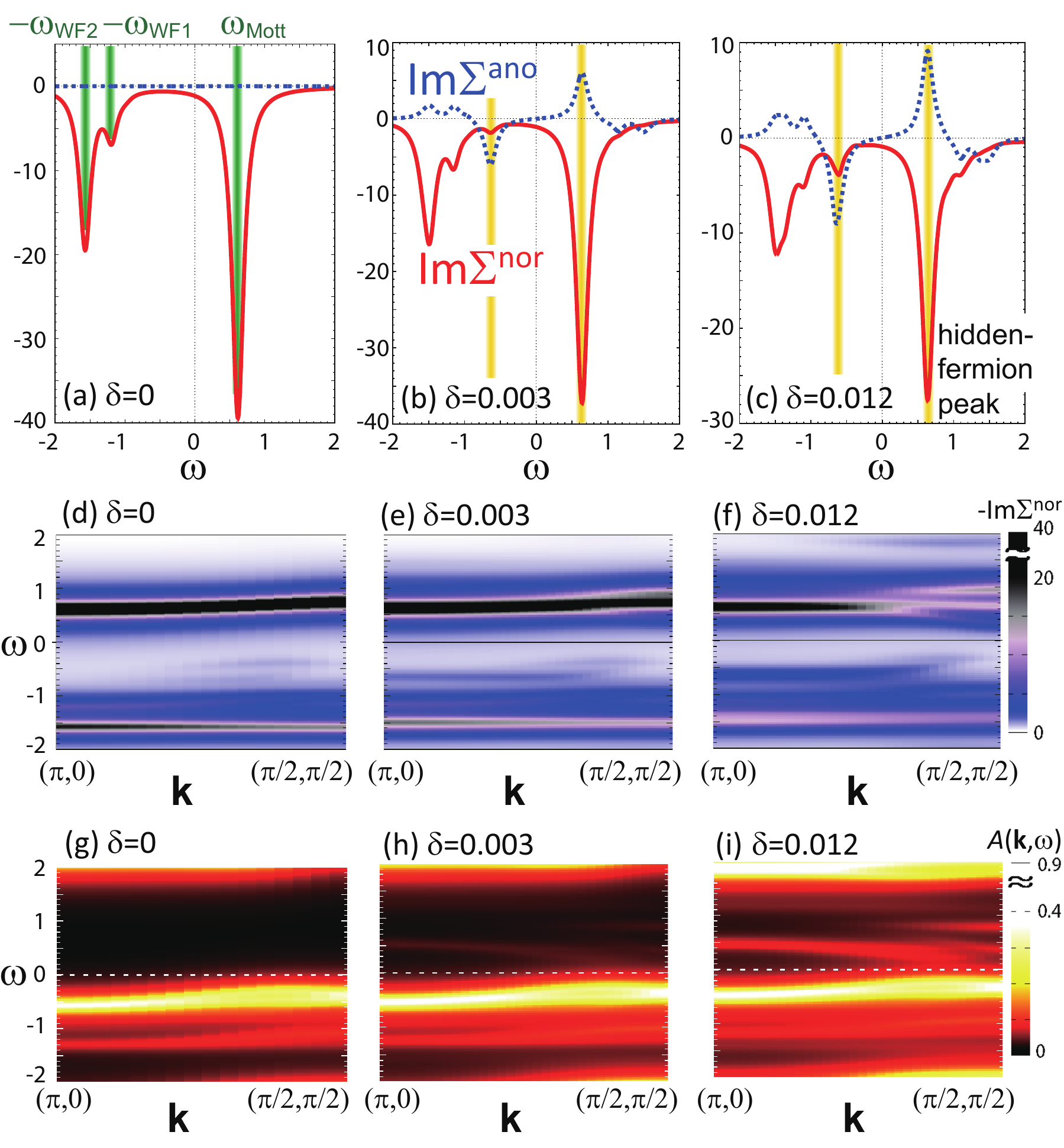}}
\caption{Doping evolution of self-energy and spectral function in the vicinity of the Mott insulator at $U=7$.
(a) Self-energy of the Mott insulator at $\d=0$. We define $\wmott$, $\wwfa$ and $\wwfb$ for this state. (b), (c) Self-energy of the superconducting state at tiny dopings. The yellow vertical lines denote the self-energy pole structure evolving into the hidden-fermion peak. (d),(e),(f) Intensity plot of -Im$\Snor$ along the $(\pi,0)-(\frac{\pi}{2},\frac{\pi}{2})$ line. 
(g),(h),(i) Corresponding plots of $A(\Vec{k},\w)$.}
\label{fig:sigma7}
\end{figure}

We start with the Mott insulating state, which is obtained for $U\gtrsim 6$ in the $2\times2$ CDMFT \cite{zhang07,park08}.
To make it simpler to compare with doped systems, the chemical potential is put just above the occupied states, where the calculated electron density $1-\d$ is more than 0.999.
We first look at the local density of states displayed in Fig.~\ref{fig:dos}(a). The gap just above the Fermi level ($\w=0$) is the Mott gap \footnote{The density of states remaining within the Mott gap is mainly due to the effect of the energy broadening $i\eta$ rather than the finite temperature effect.}. 
Above the Mott gap we see the upper Hubbard band (UHB) while the occupied states below the Mott gap is the lower Hubbard band (LHB).
The Mott gap increases as $U$ increases while the shape of each Hubbard band changes only weakly with $U$. 
The dip around $\w\simeq -1.5$ is related to the suppression of $\Akw$ in this energy range, as shown in Fig.~\ref{fig:dos}(b).
Because this behavior has been discussed in previous theoretical works \cite{macridin07,sakai10,kohno14} in connection with the waterfall  behavior observed in cuprates \cite{graf07,valla07,xie07,meevasana07}, we shall call in this paper the dip feature the "waterfall".

In Fig.~\ref{fig:phase}(b) [Fig.~\ref{fig:sigma7}(a)], Im$\Snor$ at $\Vec{k}=\kAN$ shows mainly two structures: One is the sharp peak at $\w=\wmott\simeq1.1 (0.6)$, which generates the Mott gap, and the other is a relatively broad peak for $-2\lesssim \w\lesssim -1$, which generates the waterfall mentioned above.
Just for the sake of convenience in the following discussions, we define by $-\wwfa$ and $-\wwfb$ the position of the upper and lower peaks in this structure, as illustrated in Fig.~\ref{fig:sigma7}(a).
We use these energies in order to specify the position of the waterfall structure while we avoid to discuss its fine structure, considering the finite-size feature of our calculation.
As we shall see below, $\wwfa$ and $\wwfb$ do not strongly depend on $U$ while $\wmott$ monotonically increases with $U$ in accordance with the shift of the UHB to higher energies (see Fig.~\ref{fig:dos}).

\subsection{General remarks on the doping evolution of self-energy}\label{ssec:general}
When carriers are doped into the Mott insulator, the hidden fermion (i.e., self-energy pole enhancing the superconductivity) emerges in different ways according as the magnitude relation between $\wmott$ and $\w_\text{WF1,2}$. We shall therefore discuss each case of $\wmott < \w_\text{WF1,2}$ or $\wmott > \w_\text{WF1,2}$ separately in the following.

Here and hereafter, we use the term "hidden fermion" only for the self-energy peak which shows a cancellation between Im$\Snor$ and Im$W$ in the superconducting state, because this cancellation is a unique property caused by a coupling to a fermionic excitation\cite{sakai16}. We shall explicitly show this cancellation in Fig.~\ref{fig:cancel} below.
We however extend this definition of the hidden fermion to the normal metallic (pseudogap) state as well if the isolated peak of Im$\Snor$ continuously evolves into the hidden-fermion pole through the superconducting transition.

The following discussions are focused on the low-energy electronic structure for $|\w|<2$, in which all the ingredients to discuss the origin of the hidden fermion are contained. 
On the other hand, the higher energy structure for $\w>2$ also changes with doping. In particular, a self-energy peak develops between the ingap state and the UHB. Interestingly, this self-energy peak traces back to the waterfall structure present in the UHB at $\d=0$ [see Fig.~\ref{fig:dos}(b)].
We shall discuss these points in more detail in Appendix A.

\subsection{Doping evolution of self-energy for $\wmott < \w_\text{WF1,2}$}\label{ssec:u7}
For a relatively small $U$, $\wmott$ is substantially smaller than $\w_\text{WF1,2}$ at $\Vec{k}=\kAN$, as is displayed in Fig.~\ref{fig:sigma7}(a) for $U=7$.
A tiny doping immediately lifts $\Sano$ [Fig.~\ref{fig:sigma7}(b)].
Here, the most important finding is that Im$\Sano$ develops sharp peaks at $\w=\wmott$ and its electron-hole symmetric position [see Fig.~\ref{fig:peak}(a) below, too]:  This evidences the direct transformation of the Mott insulator into the superconductor.
With further doping, the peaks of Im$\Sano$ at $\w=\pm\wmott$ become more prominent  [Fig.~\ref{fig:sigma7}(c)], evolving into the hidden-fermion peaks similar to those seen in Fig.~\ref{fig:phase}(c). Thus, the origin of the hidden fermion is identified with the Mott gap.

On the other hand, the structure around $\w_{\rm WF1,2}$ gives a broad peak with sign opposite to Im$\Sano(\kAN,\wmott)$ (i.e., hidden-fermion peak).
Note that Im$W$ and Im$\Snor$ do not cancel each other at $\w_{\rm WF1,2}$ in this case of $U=7$, as we shall show in Sec.~\ref{ssec:chara}.
As the sign of Im$\Sano(\kAN,\w_{\rm WF1,2})$ is opposite to that of the hidden-fermion peak,
the waterfall structure at $U=7$ cannot be directly connected to the hidden fermion, in contrast to the peak at $\wmott$.
We show in the next subsection that this behavior qualitatively changes at larger $U$.

One may wonder if the above correspondence between the hidden-fermion peak and $\wmott$ holds away from $\kAN$, too, because the Mott gap has in general a much larger energy scale than the pseudogap\cite{sakai09PRL,pudleiner16}. To examine this point, we plot -Im$\Snor$ along the $(\pi,0)-(\frac{\pi}{2},\frac{\pi}{2})$ cut in Figs.~\ref{fig:sigma7}(d)-(f) [data along $(0,0)-(\pi,0)-(\pi,\pi)-(0,0)$ are shown in Appendix B, where a large dispersion of $\wmott$ is apparent]. We see that around this line, which is close to the Fermi surface in the normal-state solution of doped systems, $\wmott$ is always located at low energy ($\wmott <1$), and is indeed transformed into the hidden-fermion peak at finite dopings. Note that the peak of Im$\Snor$ splits around $(\frac{\pi}{2},\frac{\pi}{2})$ in Fig.~\ref{fig:sigma7}(f).
This feature may depend on the choice of the periodization scheme \cite{kotliar01,civelli09PRB,sakai12}, as the momentum around the nodal point is not directly accessible within the solution of the $2\times2$-cluster impurity model. However, the lowest-energy branch corresponds to the $d$-wave form of the superconducting gap. 
Spectral function obtained for these self-energies are displayed in Figs.~\ref{fig:sigma7}(g)-(i), which clearly shows the Mott gap (for $0<\w\lesssim 2$) at $\d=0$ and the $d$-wave Bogoliubov bands (at $\w\simeq \pm 0.5$ for $\Vec{k}=\kAN$) at finite dopings. Spectra along other symmetry lines are displayed in Fig.~\ref{fig:skw7} in Appendix B.

\subsection{Doping evolution of self-energy for $\wmott \gtrsim \w_\text{WF1,2}$}\label{ssec:u9}

\begin{figure}[tb]
\center{
\includegraphics[width=0.48\textwidth]{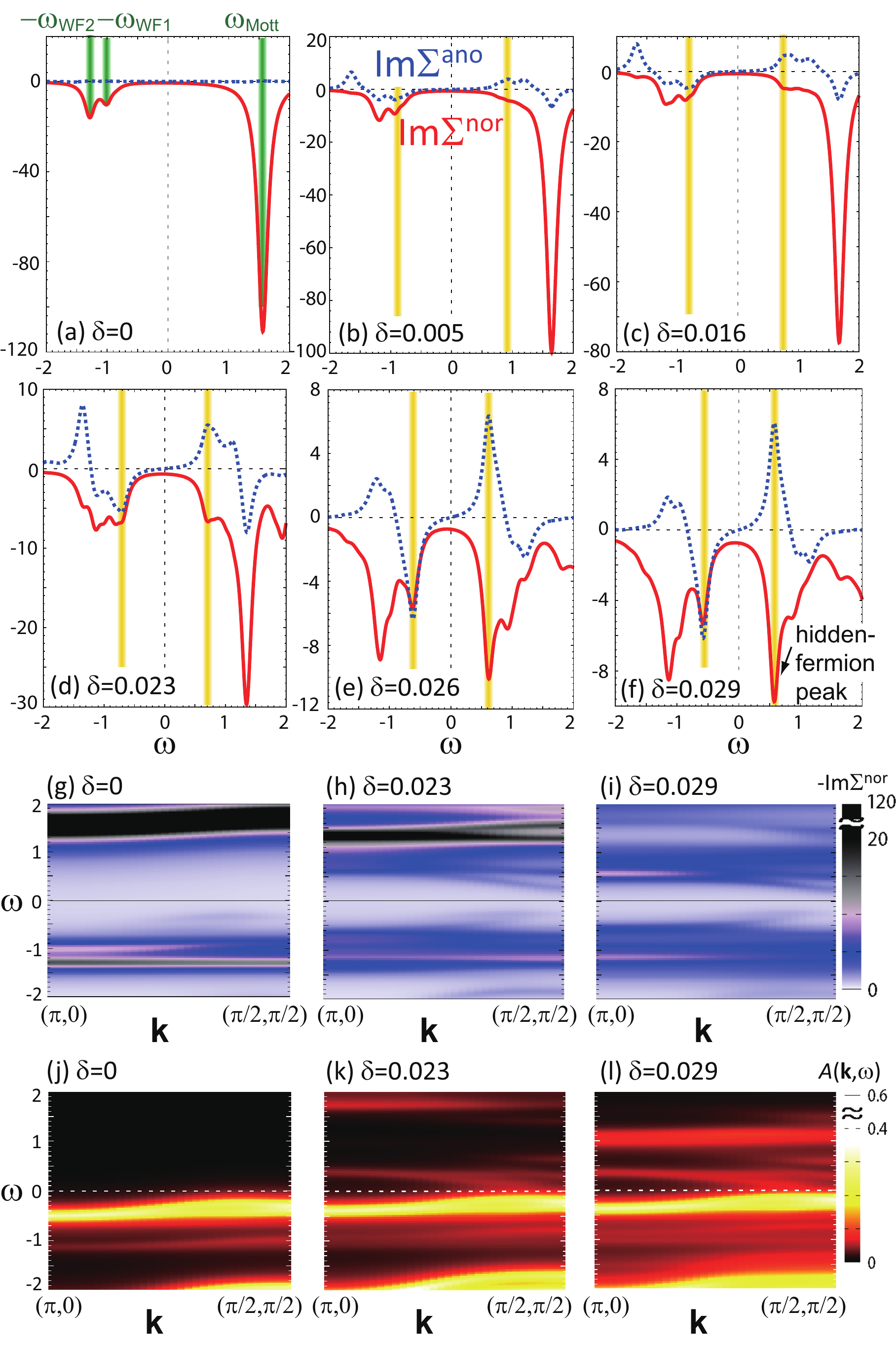}}
\caption{The same as Fig.~\ref{fig:sigma7} but at $U=9$.}
\label{fig:sigma9}
\end{figure}

For a large $U$, $\wmott$ is larger than $\w_\text{WF1,2}$ at $\Vec{k}=\kAN$.
Figure \ref{fig:sigma9} shows the results at $U=9$.
In this case, too, $\Sano$ immediately becomes finite at a tiny doping [Fig.~\ref{fig:sigma9}(b)].
However, an interesting difference from the above case of $U=7$ is that the lowest-energy peak (indicated by a yellow vertical line) of Im$\Sano$ emerges at $\w=\pm\wwfa$ [see Fig.~\ref{fig:peak}(b), too], instead of $\pm\wmott$ in the previous case. As the doping increases, this lowest-energy peak evolves into the hidden-fermion peak enhancing the superconductivity [Fig.~\ref{fig:sigma9}(f)].

In more detail, Im$\Sano$ in Fig.~\ref{fig:sigma9}(b) shows another peak at $\w=\pm\wwfb$ and an opposite-sign peak at $\w=\pm\wmott$. Unlike the case for $U=7$, the latter does not develop much for further dopings, remaining a weak opposite-sign weight above the hidden-fermion peak energy.
The energy of the hidden-fermion peak gradually decreases with doping.
In response to the change of $\Sano$, $\Snor$ also changes with doping: The Mott-gap peak  at $\w=\wmott$ rapidly loses its weight, which is partly transferred to the hidden-fermion peak.
This may be more clearly seen in Figs.~\ref{fig:sigma9}(g)-(i) and Figs.~\ref{fig:skw9}(a)-(c) in Appendix B.

In terms of the spectral function, a spectral weight descends from the UHB with doping [Figs.~\ref{fig:sigma9}(j)-(l) and Fig.~\ref{fig:gl9} in Appendix A]. A part of this weight makes up the upper Bogoliubov band while a substantial weight remains above the Bogoliubov band separated by a dip of the spectral weight. 
Note that the split between the upper Bogoliubov band and the weight above it can be seen in Fig.~\ref{fig:sigma7}(i), too. This split has been observed in electronic Raman scattering experiments for various cuprates as a peak-dip feature in the $B_{1g}$ response \cite{loret16,loret17} and comes from the pole cancellation discussed in Sec.~\ref{ssec:chara}.

For $U=8$, $\wmott$ is comparable to $\w_\text{WF1,2}$ at $\Vec{k}=\kAN$. In this case, both $\wmott$ and $\w_\text{WF1,2}$ are involved in the emergence of the hidden fermion peak at a tiny doping so that it requires a more careful analysis. We discuss this case in Appendix C.

\subsection{Relationship between Mott insulator and high-$\Tc$ superconductivity}

\begin{figure}[tb]
\center{
\includegraphics[width=0.48\textwidth]{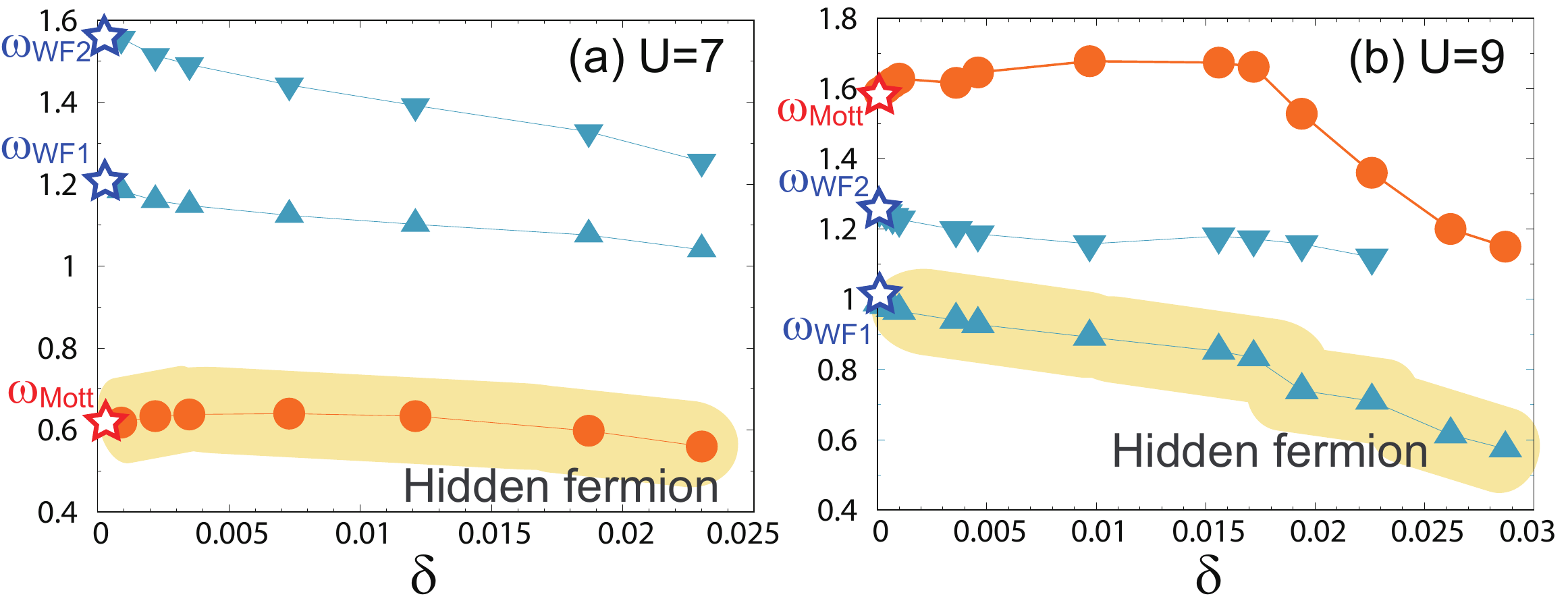}}
\caption{Peak position of the self-energy plotted against doping $\d$ for (a) $U=7$ and (b) $U=9$. Stars denote $\wmott$ and $\w_\text{WF1,2}$ measured with Im$\Snor(\kAN,\w)$ of the Mott insulator. Orange circles and light-blue triangles denote the peak positions of Im$\Sano$ in the superconducting state, where the former and latter have the intensity of opposite signs. The hidden fermion peak is hatched by yellow. In panel (b), the inverted triangle is not plotted for $\d>0.025$, where the peak is not discernible likely because of the proximity to another peak denoted by orange circles. }
\label{fig:peak}
\end{figure}

From Figs.~\ref{fig:sigma7} and \ref{fig:sigma9} (and additional data), we have extracted the peak positions of the self-energy, and plotted them against $\d$ in Fig.~\ref{fig:peak}. 
The star symbols denote $\wmott$ and $\w_\text{WF1,2}$ extracted from Im$\Snor$ in the Mott insulator. 
We see $\wmott<\wwfa$ ($\wmott>\wwfa$) for $U=7$ ($U=9$).
Circles and triangles plot the peak positions of Im$\Sano$ in doped cases \footnote{We have used $\Sano$ rather than $\Snor$ because the peaks are more discernible for the former, though in most cases both show peaks at the same energy.}, where the former and the latter denote the opposite-sign peaks.  
These plots clearly show the continuity of the peak positions, and in particular the origin of the hidden-fermion peak, which develops as the lowest-energy peak for $\d>0$;
for $U=7$, it is $\wmott$, and for $U=9$ it is $\wwfa$.

\begin{figure}[tb]
\center{
\includegraphics[width=0.48\textwidth]{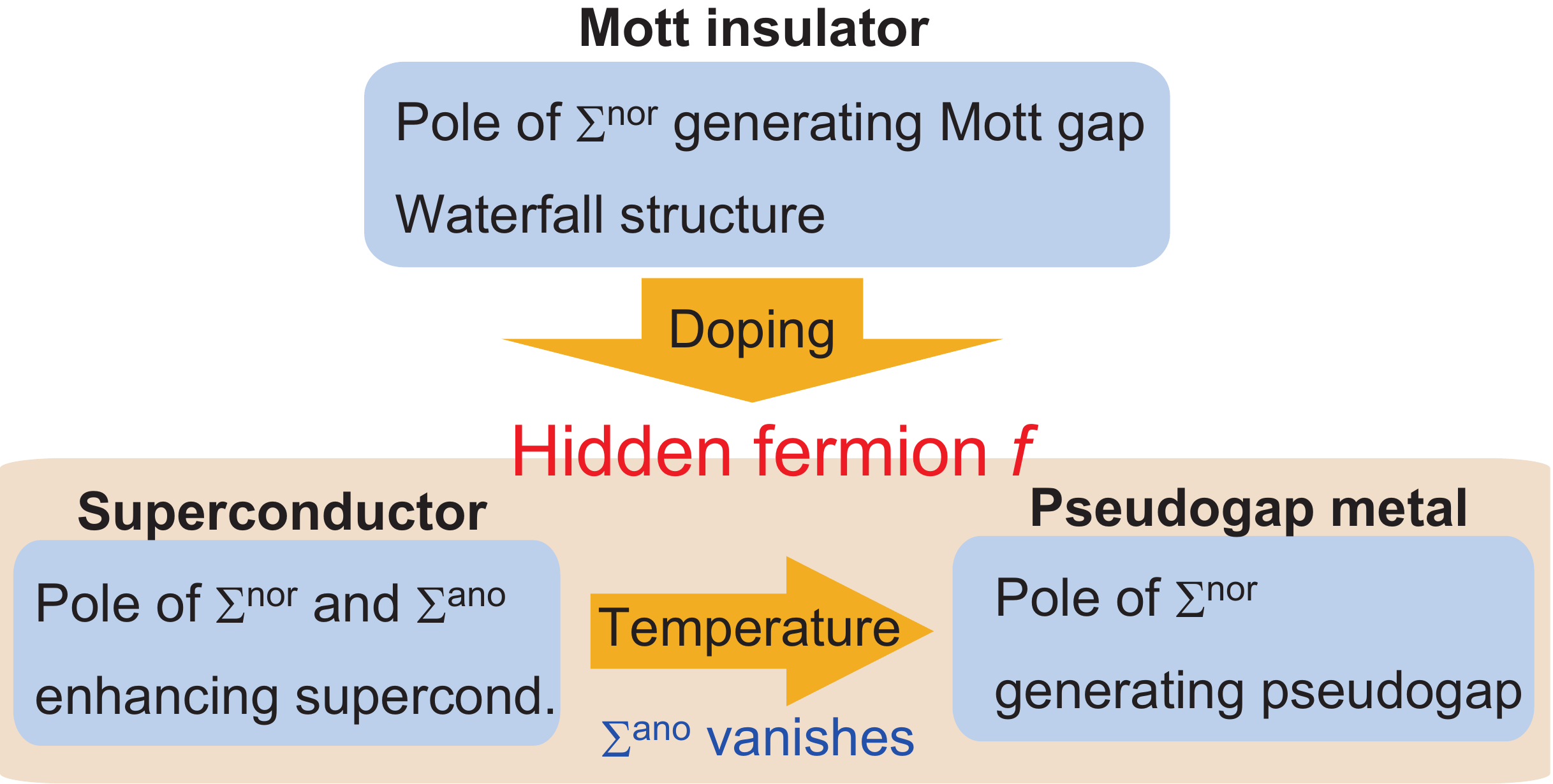}}
\caption{Relation between the self-energy structures in the Mott insulator, pseudogap metal, and superconducting state, inferred from the present results and those in Refs.~\onlinecite{sakai16,sakai09PRL}.}
\label{fig:relation}
\end{figure}

We have thus established the microscopic relation between the Mott insulator and the high-$\Tc$ superconductivity: The hidden-fermion peak enhancing the superconductivity traces back to the self-energy peaks present in the Mott insulator.
For $\wmott<\w_\text{WF1,2}$ it traces back to the Mott-gap peak, while for $\wmott > \w_\text{WF1,2}$ it does to the waterfall peak (Fig.~\ref{fig:relation}). 
The former is the case for a relatively small $U$ ($6\lesssim U \lesssim 7$) while the latter is the case for a relatively large $U$ ($\gtrsim 8$).
This amplitude relation between $\wmott$ and $\w_\text{WF1,2}$ may change with the momentum, too (Appendix B).
We emphasize that no symmetry breaking, like antiferromagnetism or charge order, has a direct relevance to the above mechanism generating the hidden fermion.

As we show in Appendix D, for an energetically isolated pole of the self-energy, its residue of the normal component ($R_\pm^\text{nor}$ for the poles at the positive/negative energies, respectively) and that of the anomalous component ($R_\pm^\text{ano}$) satisfy\begin{align}
R_{+}^\text{nor}R_{-}^\text{nor}=(R_{\pm}^\text{ano})^2,
\end{align} 
where $R_+^\text{ano}=-R_-^\text{ano}$ holds. 
This relation explains the above evolution of the self-energy peaks in some more detail.
For example, in Fig.~\ref{fig:sigma7}(a) to (c), as doping increases, $R_+^\text{nor}$ decreases while $R_-^\text{nor}$ increases in a way that their product, $R_{+}^\text{nor}R_{-}^\text{nor}$, increases. 
Accordingly, $|R_\pm^\text{ano}|$ also increases. 
Thus, the relation between $R_\pm^\text{nor}$ and $R_\pm^\text{ano}$ inferred in the above equation supports the picture that the large normal self-energy present in the Mott insulator is the source of the large $\Sano$ in the superconductors. 

The results in Ref.~\onlinecite{sakai16PRB} suggest that the superconductivity is maximized around $U=7-8$. This fact may also be explained by the picture obtained above. Namely, the availability of the Mott-gap pole at low energy and its strength increasing with $U$ may have a good balance around $U=7-8$, leading to a strong superconductivity.

In Appendix E, we have given a more detailed analysis of the doping evolution of the self-energy. The analysis demonstrates that the difference between $U=7$ and $U=9$ cases can indeed be ascribed to the difference of the amplitude relation between $\wmott$ and $\wwfa$;  the doping dependence of other ingredients, like the weight of the self-energy peaks and the strength of the anomalous part, is shown to be qualitatively similar in both cases. 

The above discussions concern the low-energy structure of the self-energy while in Appendix A we present a higher-energy structure and discuss its doping evolution. In fact, the weight of the Mott-gap peak at $\d=0$ is transferred, with doping, to a higher-energy structure, too, which then makes a gap between the ingap state and UHB, according to the sum rule of the self-energy weight in the superconducting state \cite{sakai16PRB}. This behavior is observed generally for $U=7,8$ and $9$.

\subsection{Characterization of self-energy peaks}
\label{ssec:chara}
 
The broad self-energy peak structure around  $-\w_\text{WF1,2}$ yields a suppression of the spectral function. 
This suppression has been found at $\d=0$ in previous numerical studies \cite{kyung06PRB1,civelli09PRB,sakai09PRL,liebsch09,sakai10,kohno12,kohno14,pudleiner16},
and found to persist in the normal-state solution at finite dopings \cite{maier02,senechal04,macridin07, kyung06PRB1,civelli09PRB,sakai09PRL,liebsch09,sakai10,kohno12,kohno14}.
In Refs.~\onlinecite{macridin07,sakai10,kohno14}, this structure has been discussed in connection with the high-energy kink or waterfall structure observed in the cuprates \cite{graf07,valla07,xie07,meevasana07}.
A similar structure has been seen even within the single-site DMFT calculations \cite{karski05,byczuk07,karski08,raas09,lee17}, which takes into account only local correlations, as well as in angle-resolved photoemission spectra of SrVO$_3$ \cite{aizaki12,yoshida16}.
These observations indicate that this structure is a direct consequence of the Mott physics, irrespective of the spatial dimensions and lattice structures.

In order to elucidate the origin of the self-energy peaks, we investigate whether the peaks of ${\rm Im}\Snor$ and ${\rm Im}W$ [Eq.~(\ref{eq:W})] cancel with each other: A cancellation signifies the isolated pole character of the peak while the absence of the cancellation signifies a continuous spectrum of the excitation.

\begin{figure}[tb]
\center{
\includegraphics[width=0.48\textwidth]{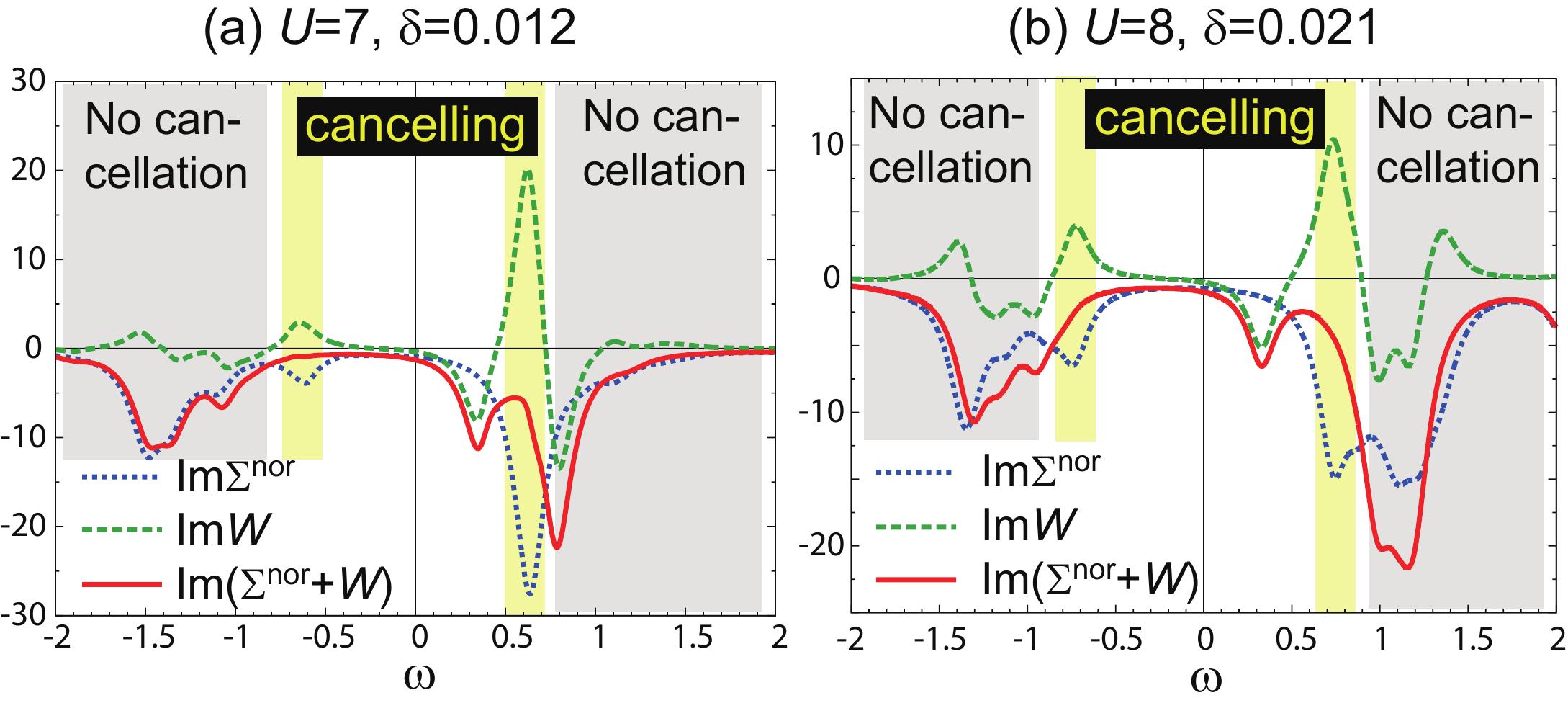}}
\caption{Relation between the normal and anomalous contributions to Green's function in two slightly doped cases. In the yellow region, Im$\Snor$ and Im$W$ have opposite signs, and the peaks cancel out in their sum. In the gray region, Im$W$ has the same sign as Im$\Snor$ or is so small that their sum follows the curve of Im$\Snor$.}
\label{fig:cancel}
\end{figure}

Figure \ref{fig:cancel} plots the imaginary part of $\Snor$, $W$, and their sum at tiny dopings. The hidden-fermion peaks are discernible as negative-intensity peaks in Im$\Snor$ (see yellow area), and at the same energies, Im$W$ shows positive-intensity peaks, which result from the peak of Im$\Sano$ at these energies.
In their sum, Im$(\Snor+W)$, however, no trace of the peak is discernible. Namely, the peak weights of Im$\Snor$ and Im$W$ cancel out. In Ref.~\onlinecite{sakai16}, we revealed that this cancellation is a direct consequence of a {\it fermionic} excitation yielding an isolated pole in the self-energy.

On the other hand, in the higher-energy region colored by gray, Im$W$ shows negative or small positive values, so that no cancellation occurs and Im$(\Snor+W)$ more or less follows the curve of Im$\Snor$.
This means that the broad self-energy peaks around $\w=\pm\w_\text{WF1,2}$ cannot be described by energetically-isolated fermionic excitations but will be described by a continuum of them \cite{sakai16PRB} and perhaps be effectively described by a coupling to  a bosonic excitation.

\begin{figure}[tb]
\center{
\includegraphics[width=0.48\textwidth]{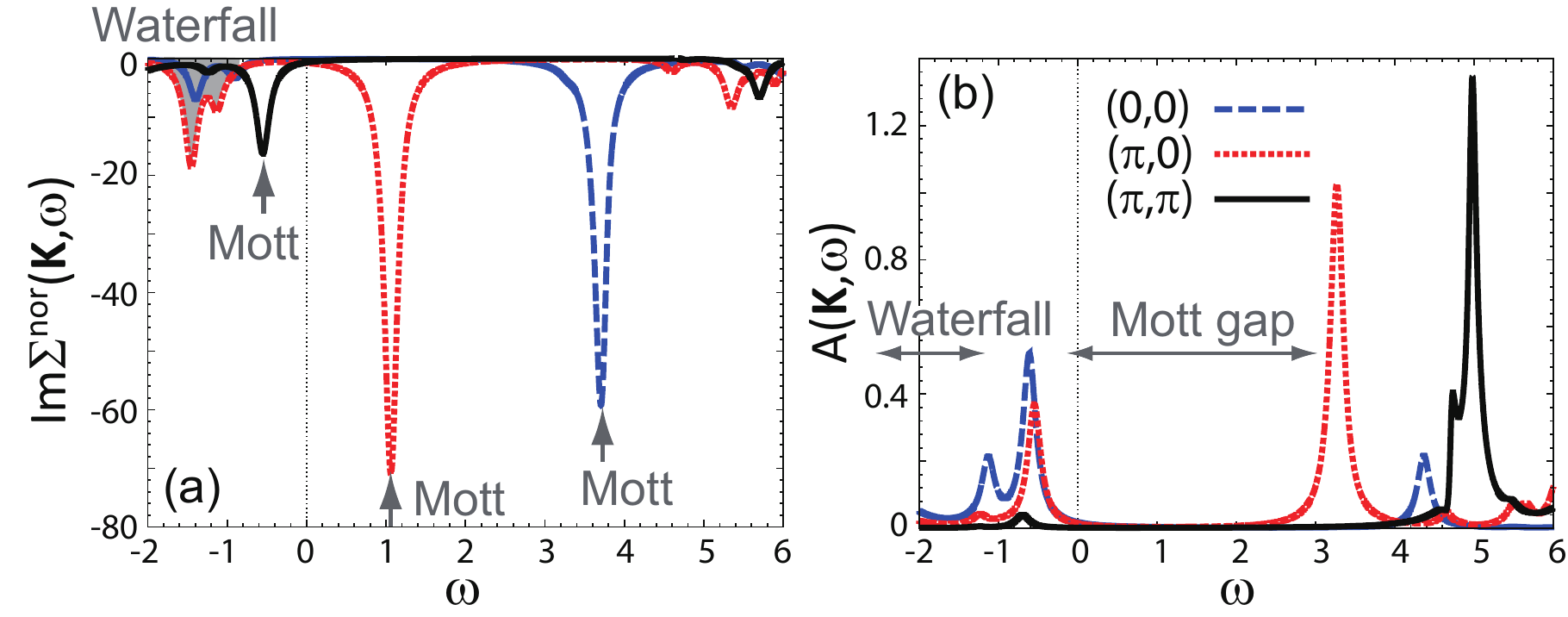}}
\caption{(a) Im$\Snor(\Vec{K},\w)$ and (b) $A(\Vec{K},\w)$ at cluster momenta $\Vec{K}=(0,0),(\pi,0)$ and $(\pi,\pi)$, calculated for the Mott insulating state at $\d=0$ and $U=8$. Arrows in panel (a) indicate the self-energy peak generating the Mott gap while the shaded area denotes the one generating the waterfall.}
\label{fig:sigmaK}
\end{figure}

In order to further elucidate the character of these excitations, we plot in Fig.~\ref{fig:sigmaK}(a) Im$\Snor$ at cluster momenta $\Vec{K}=(0,0),(\pi,0)$ and $(\pi,\pi)$ for the Mott insulator.
The result reveals that the self-energy peak generating the Mott gap is dispersive \cite{sakai09PRL,pudleiner16}
while that of the waterfall is not.
The latter indicates that this excitation is spatially localized.

\section{Discussion}
\label{sec:discuss}

\subsection{Interpretation of general self-energy structure}

In the following, we discuss a possible interpretation of the above numerical results.
The interpretation is based on the observation in our previous work \cite{sakai15,sakai16,sakai16PRB} that the self-energy peaks can be represented by auxiliary {\it fermionic} degrees of freedom $f_\a$'s hybridizing with a bare electron (or a low-energy electron when we focus on the low-energy electronic structure) $c$. 
These auxiliary fermionic degrees of freedom represent correlated electronic states, to and from which the bare electron transits; this process gives the frequency-dependent self-energy.
Note that, while this multiple-auxiliary-fermion description of the correlation effect is always possible, in this article we use the term "hidden fermion" to point at a specific excitation which appears in the superconducting state (and in the pseudogap state above $\Tc$) as an isolated pole in the low-energy part of the self-energy.

For instance, in the normal state, we consider the following effective Hamiltonian,
\begin{align}
H_\text{eff}=&\sum_{\Vec{k}\s} \left[\e(\Vec{k})+\frac{U}{2}(1-\d)\right] c_{\Vec{k}\s}^\dagger c_{\Vec{k}\s} \nonumber\\
+&\sum_{\a\Vec{k}\s}  \left[\e_{f_\a}(\Vec{k}) f_{\a\Vec{k}\s}^\dagger f_{\a\Vec{k}\s} + V_\a(\Vec{k}) (c_{\Vec{k}\s}^\dagger f_{\a\Vec{k}\s} + \text{h.c.}) \right].
\label{eq:heff}
\end{align}
Integrating out the $f$ degrees of freedom in the corresponding action, we obtain \cite{sakai16PRB}
\begin{align}
\Snor(\Vec{k},\w)=\frac{U}{2}(1-\d)+\sum_\a \frac{V_\a(\Vec{k})^2}{\w-\e_{f_\a}(\Vec{k})}
\end{align}
as the self-energy of the electron $c_{\Vec{k}\s}$.
Continuous distribution of $\e_{f_\a}$ can represent a broad peak of Im$\Snor$. 
On the other hand, when there is an $\e_{f_\a}$ energetically isolated from other $\e_{f_\a}$'s, it represents a self-energy pole. This self-energy pole splits $\Akw$ into two parts below and above $\e_{f_\a}$, which can be interpreted as a bonding/antibonding state of $c$ and $f_\a$.

\begin{figure}[tb]
\center{
\includegraphics[width=0.48\textwidth]{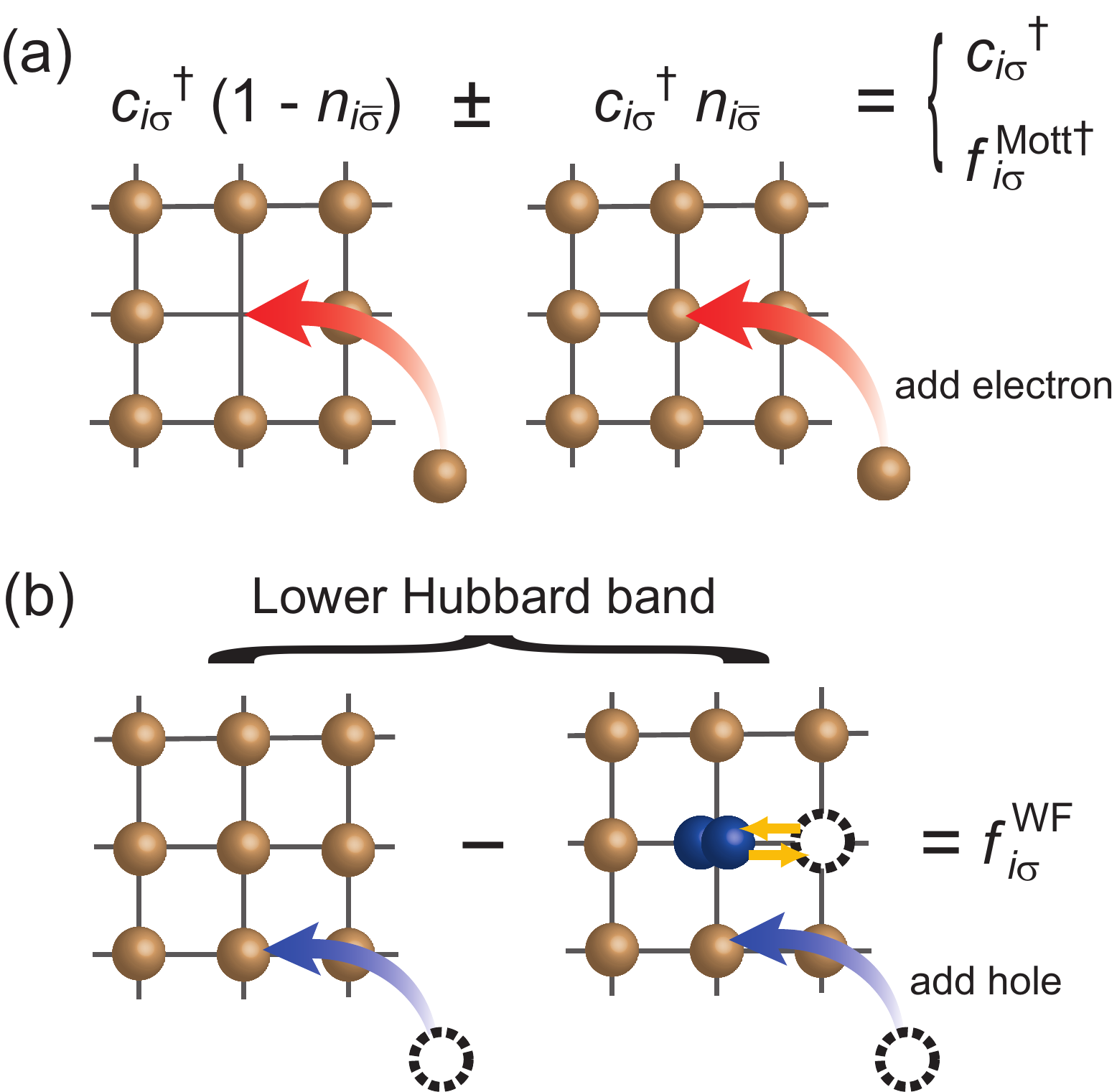}}
\caption{Schematic illustration of the fermionic excitations in the Mott insulator at $\d=0$.  Small khaki circle represents an electron. (a) Those relevant to the Mott gap, where the effect of $t$ is not illustrated explicitly for simplicity. ${\fmott}^\dagger$ represents the electron addition to the antibonding combination of the empty and singly-occupied states. (b) Internal structure of the occupied states (LHB) in the presence of a finite hopping $t$. $\fwf$ represents the hole addition to the antibonding combination of the two states with and without the dynamical doublon-hole excitations.}
\label{fig:mott}
\end{figure}

\subsection{Interpretation of self-energy structure in Mott insulator}

In the Mott-insulating state at $\d=0$, the self-energy pole generating the Mott gap splits the spectrum into the UHB and LHB.
As the UHB (LHB) basically represents doubly (singly) occupied states, this self-energy pole represents a linear combination of singly and doubly occupied states.
In fact, in the atomic limit ($t=0$), this superposed fermionic state is represented by $f_{i\sigma}^{\text{Mott} \dagger} \equiv c^\dagger_{i\sigma}(1-2n_{i\bar{\sigma}})=c^\dagger_{i\sigma}(-1)^{n_{i\bar{\sigma}}}$ [Fig.~\ref{fig:mott}(a)]\cite{zhu13,imada18}, which we call the Mott fermion. 
Then, the hybridization between $c$ and $f^\text{Mott}$ gives the LHB and UHB as the bonding and antibonding states, i.e., $c^\dagger_{i\s}+f_{i\s}^{\text{Mott} \dagger} = c^\dagger_{i\s}(1-n_{i\bar{\sigma}})$ and $c^\dagger_{i\s}-f_{i\s}^{\text{Mott} \dagger} = 2c^\dagger_{i\s}n_{i\bar{\s}}$, respectively.

For finite $t$, the Mott fermion acquires a mobility. 
The doublon in the Mott insulator can have a rather large mobility though it is somewhat suppressed compared to the bare bandwidth due to the renormalization by the antiferromagnetic fluctuations. This intuitively explains a large dispersion (as large as $3t$) of the Mott fermion seen in Fig.~\ref{fig:sigmaK}(a) and panel (a) of Figs.~\ref{fig:skw7}, \ref{fig:skw8} and \ref{fig:skw9} in Appendix B.
Under a strong antiferromagnetic correlation present in the Mott insulator, the factor $(-1)^{n_{i\bar{\sigma}}}$ gives $f_{i\sigma}^{\text{Mott} \dagger}$ a nearly $(\pi,\pi)$-displaced dispersion compared to that of electrons \cite{zhu13}.
This $(\pi,\pi)$-displaced dispersion of the Mott fermion (with a reduced bandwidth mentioned above) is indeed seen in panel (a) of Figs.~\ref{fig:skw7}, \ref{fig:skw8} and \ref{fig:skw9} in Appendix B, where the bottom and the top of the Mott-gap peak are located at $(\pi,\pi)$ and $(0,0)$, respectively.

Since the Mott fermion is a fermion in the resonating state of the UHB and LHB, it is interpreted as a fermion added to a resonating doublon-hole pair. This pair is an exciton in the Mott insulator for $t\ne 0$.
Namely, one can add a local Mott fermion only at the site represented by the linear combination of the electron-empty and singly occupied state, because after adding the Mott fermion, the state becomes the linear combination of singly and doubly occupied states. Such a resonating state with empty and singly-occupied sites is nothing but the dynamical exciton state where an exciton $(0,2)$ and a singly-occupied pair $(1,1)$ are resonating in the notation $(n,m)$ for $n$ and $m$ electrons at the neighboring sites. This means that the Mott fermion resides in the underlying vacuum fluctuation generating the exciton.

Finite $t$ also generates doublon-hole pairs dynamically in the Mott insulator.
This creates an internal structure in each Hubbard band. 
As for the LHB, electronic states involving dynamical doublon-hole pairs should have a relatively high energy among the occupied states. Therefore, they are located close to the top of LHB. 
On the other hand, the states close to the bottom of LHB will be well described by a simple singly-occupied state.
A hole is added to either of these two states (with or without the dynamical doublon-hole pairs) when we look at the occupied spectra.
Then, in the same way as above, the antibonding combination of the two different hole operators (projected onto the above two different states) will give $f_{i\sigma}^{\text{WF} }$ [Fig.~\ref{fig:mott}(b)], which represents the self-energy peak generating the waterfall structure.
Namely, the hybridization between $c$ and $\fwf$ produces the spectral weights below and above the waterfall structure, as the bonding and antibonding states, respectively.
Here, $f_{i\sigma}^{\text{WF}}$ will have a continuous spectrum, as indicated in Fig.~\ref{fig:cancel}, because various dynamical doublon-hole excitations may be considered. 
The observation in Fig.~\ref{fig:sigmaK} that $f_{i\sigma}^{\text{WF}}$ is localized in space is also compatible with this picture because the doublon-hole pairs dynamically generated in the Mott insulator can hardly move around.

\begin{figure}[tb]
\center{
\includegraphics[width=0.48\textwidth]{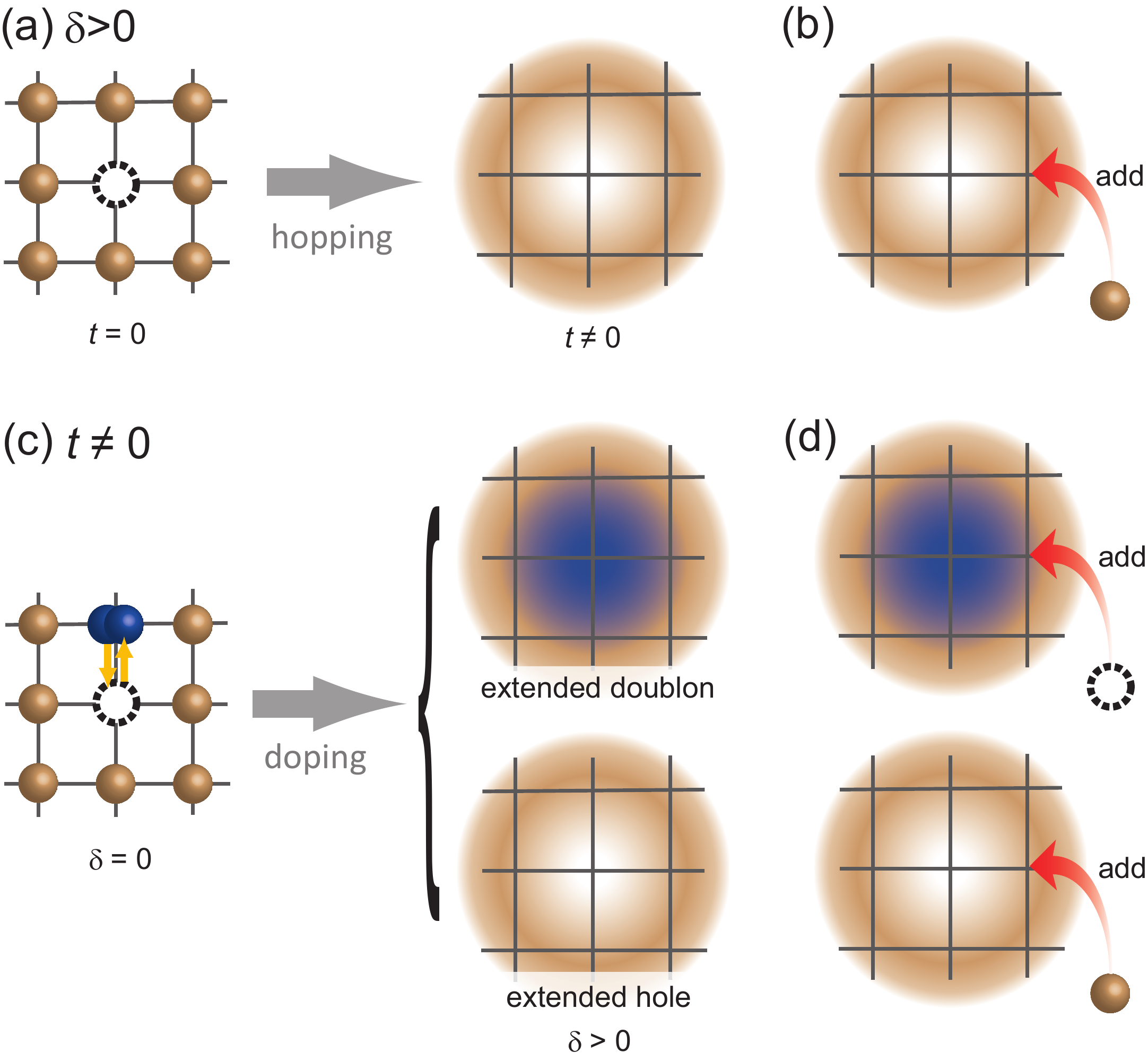}}
\caption{Schematic illustration of electronic states relevant to the hidden fermion excitation. Broad khaki area represents a delocalized electron. (a) Doped holes (dashed open circle) get delocalized when the hopping $t$ is switched on. (b) An electron added around this hole will be weakly bound to the hole. We speculate that this electron bound to the hole comprises the hidden fermion. (c) The dynamical doublon-hole pairs present at $\d=0$ is delocalized due to hole doping. The doublon (hole) part is illustrated by blue (white) in the upper (lower) panel. (d) A hole (electron) added around the doublon (hole) is bound to the doublon (hole). As panel (b) does, the lower panel represents the hidden fermion (b) while the upper one represents its hole counterpart.}
\label{fig:hole}
\end{figure}

\subsection{Interpretation of self-energy structure in doped Mott insulator}

When the system is doped with holes, the finite $t$ gives a mobility to the electrons and holes, making them delocalized.
Namely, the wave functions of electrons and holes become extended in real space [Fig.~\ref{fig:hole}(a)]. Then, an electron added around this extended hole can be weakly bound to this hole [Fig.~\ref{fig:hole}(b)].
Here, we consider a weak binding, rather than a strong binding (in the energy scale of $U$) as in the Mott insulator, because the attraction between the electron and the hole is screened by other doped holes.
This electron weakly bound to a hole is an excitonic bound state proposed in Refs.~\onlinecite{imada11,yamaji11PRL} and identified with the hidden fermion discussed in Ref.~\onlinecite{sakai16} after considering the antibonding combination with $c$ as is done above.
Since in the limit $\delta\rightarrow 0$, this hidden-fermion excitation reduces to the Mott fermion $f^{\text{Mott} \dagger}$ \footnote{To see the correspondence of the hidden fermion to $f^{\text{Mott} \dagger}$ in the limit $\delta\rightarrow 0$, a momentum-space picture would be more appropriate because the correspondence is seen only in a limited area (e.g., around $(\pi,0)-(0,\pi)$ line) of the momentum space (Appendix B).}, in this interpretation it is obvious that the hidden fermion emerges from $\w=\pm\wmott$ at tiny doping.
In other words, the Mott fermion and the hidden fermion are essentially the same in the limit $\delta\to 0$ in the momentum region where $\wmott <\w_{\rm WF1,2}$ is satisfied.
It is remarkable that the same fermionic excitation induces very different phenomena depending on doping concentration, i.e., the Mott insulator at $\d=0$ and the high-$\Tc$ superconductivity at $\d>0$.
In Sec.~\ref{ssec:general}, we have extended the terminology of "hidden fermion" even for the isolated self-energy pole in the normal metallic phase, if the pole evolves continuously into the hidden fermion in the superconducting state. Similarly, one can regard the Mott fermion as belonging to the same category of the "hidden fermion" when it continuously evolves into the hidden fermion in the superconducting state. However, even in this case, we do not use the name of "hidden fermion" for the Mott fermion by emphasizing its special role in the Mott insulator.

In the occupied state, dynamically generated doublons and holes are also delocalized owing to the hole doping [Fig.~\ref{fig:hole}(c)].
A hole added around this extended doublon can be weakly bound to this doublon [Fig.~\ref{fig:hole}(d) upper panel]. This is the hole-type excitation of the hidden fermion discussed above. At the same time, the delocalization of the doublon-hole pairs (in other words, emergence of unbound doublon and hole) allows an electron addition near the hole to form a weakly bound pair [Fig.~\ref{fig:hole}(d) lower panel]. This gives the particle counterpart of the above hole-type hidden fermion and is nothing but the hidden fermion of Fig.~\ref{fig:hole}(b).
Because the extended doublon-hole pair is continuously connected to the dynamical doublon-hole pair (which is at the origin of the waterfall) in the Mott insulator in the limit $\d\rightarrow 0$, the hidden fermion can emerge from $\w=\pm\w_\text{WF}$ at tiny doping.
Since the lower energy excitation will be more stable, the lower one between $\wmott$ and $\w_\text{WF1}$ would determine the energy from which the hidden fermion first appears at a tiny doping.

Provided that the hidden fermion is an electron bound to a hole, it would have a dipole moment. Then, the dipole-dipole interaction would play a role of the pairing interaction  between the hidden fermions\cite{imada11}.
Through the hybridization, this pairing of the hidden fermions considerably enhances the  pairing of quasiparticles\cite{sakai16}.

\section{Conclusion}

In summary, we have presented a microscopic relationship between the Mott insulator and the high-temperature superconductivity in terms of the self-energy structure. The revealed direct relationship between the two self-energy structures explains why the superconductivity can have a high $\Tc$ in the vicinity of the Mott insulator. In short, a large self-energy present in the Mott insulator is directly transformed, with doping, into a self-energy pole  of the hidden fermion, which in turn enhances the superconductivity.

We have shown a continuous evolution of the self-energy from the Mott insulator to the superconductor, by studying an extremely small doping region. The numerical result shows that the hidden-fermion peak enhancing the superconductivity, as well as generating the pseudogap above $\Tc$, traces back to either the self-energy pole generating the Mott gap or a broader self-energy peak generating the waterfall structure at $\d=0$.
This mechanism does not rely on any specific fluctuations but is a direct consequence of the Mott physics.

The detail of this self-energy evolution depends on the value of $U$, or more explicitly the magnitude relation between $\wmott$ and $\w_\text{WF1}$ at $\d=0$: The one at the energy closer to the Fermi level seems to determine the energy from which the hidden fermion is born at a tiny doping. The magnitude relation can also change with momentum because $\wmott$  is much more dispersive than $\w_\text{WF1,2}$.
The Mott-gap and waterfall peaks of self-energy can play a similar role presumably because both accompany a doublon bound to a hole, from which the hidden fermion emerges at a finite doping.

\begin{acknowledgments}
S.S. thanks A. Liebsch for useful discussions in developing the numerical simulation code used in the present study.
S.S. is supported by JSPS KAKENHI (Grant No. JP17K14350 and JP16H06345).
M.I. is supported by JSPS KAKENHI (Grant No. JP16H06345) and by MEXT as a social and scientific priority issue (Creation of new functional devices and high-performance materials to support next-generation industries CDMSI) to be tackled by using post-K computer, and RIKEN Advanced Institute for Computational Science (AICS) through
HPCI System Research Project (Grants No. hp150211, No. hp160201 and No. hp170263), from MEXT, Japan.
\end{acknowledgments}

\section*{Appendix A: Electronic structure in a wide energy range}
\begin{figure}[tb]
\center{
\includegraphics[width=0.48\textwidth]{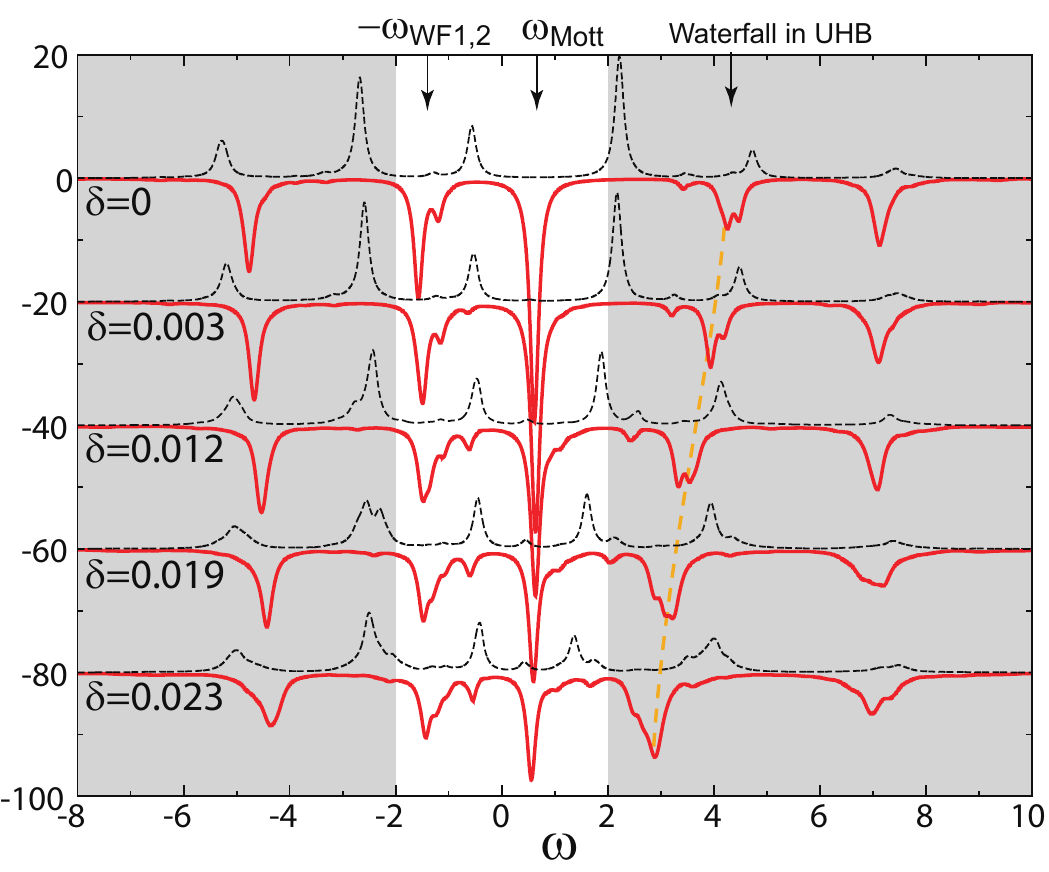}}
\caption{Im$\Snor$ (red solid curve) and $A(\Vec{k},\w)$ (black dashed curve) at $\Vec{k}=\kAN$ for $U=7$. Each curve is shifted by -20 along the vertical axis. $A(\Vec{k},\w)$ is amplified with a factor of 20. We shaded the high-energy area which has not been the focus of the present paper. The orange dashed curve indicates the self-energy peak which develops with doping from a waterfall peak in UHB at $\d=0$ to a peak giving a large gap between the ingap state and UHB at finite dopings. }
\label{fig:gl7}
\end{figure}

\begin{figure}[tb]
\center{
\includegraphics[width=0.48\textwidth]{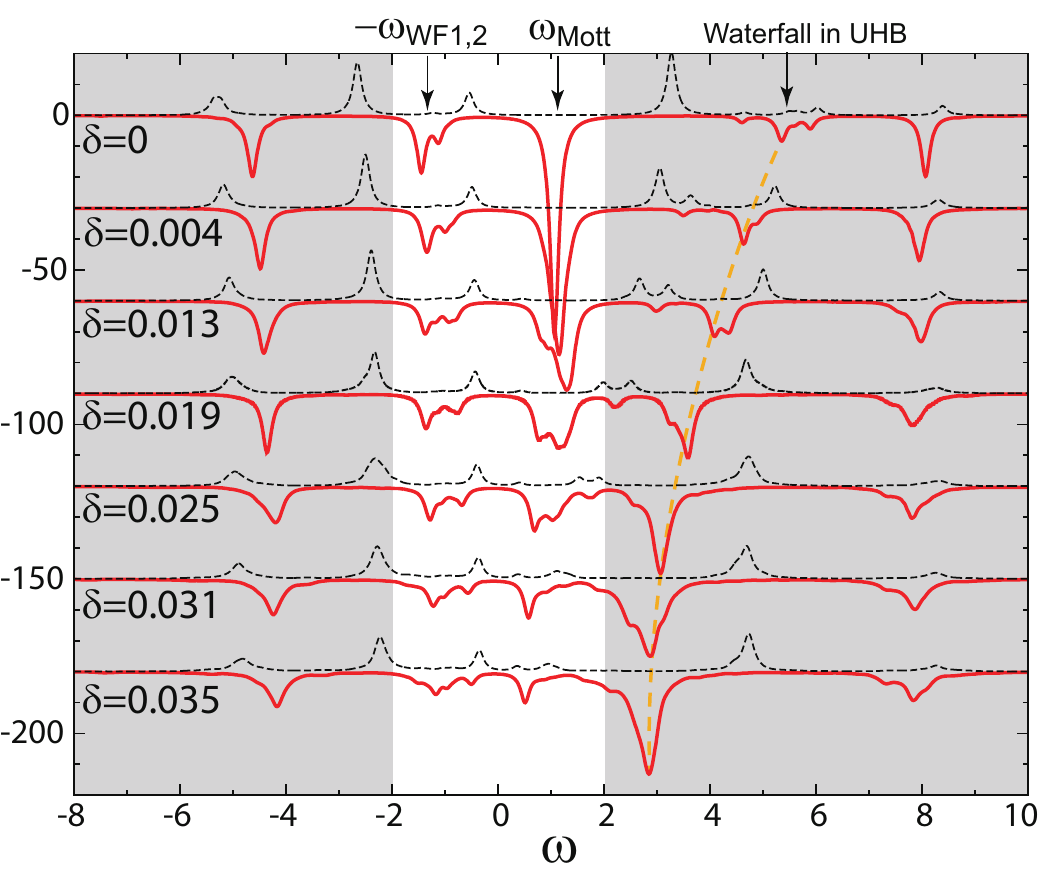}}
\caption{Same as Fig.~\ref{fig:gl7} but for $U=8$. Each curve is shifted by -30 along the vertical axis.}
\label{fig:gl8}
\end{figure}

\begin{figure}[tb]
\center{
\includegraphics[width=0.48\textwidth]{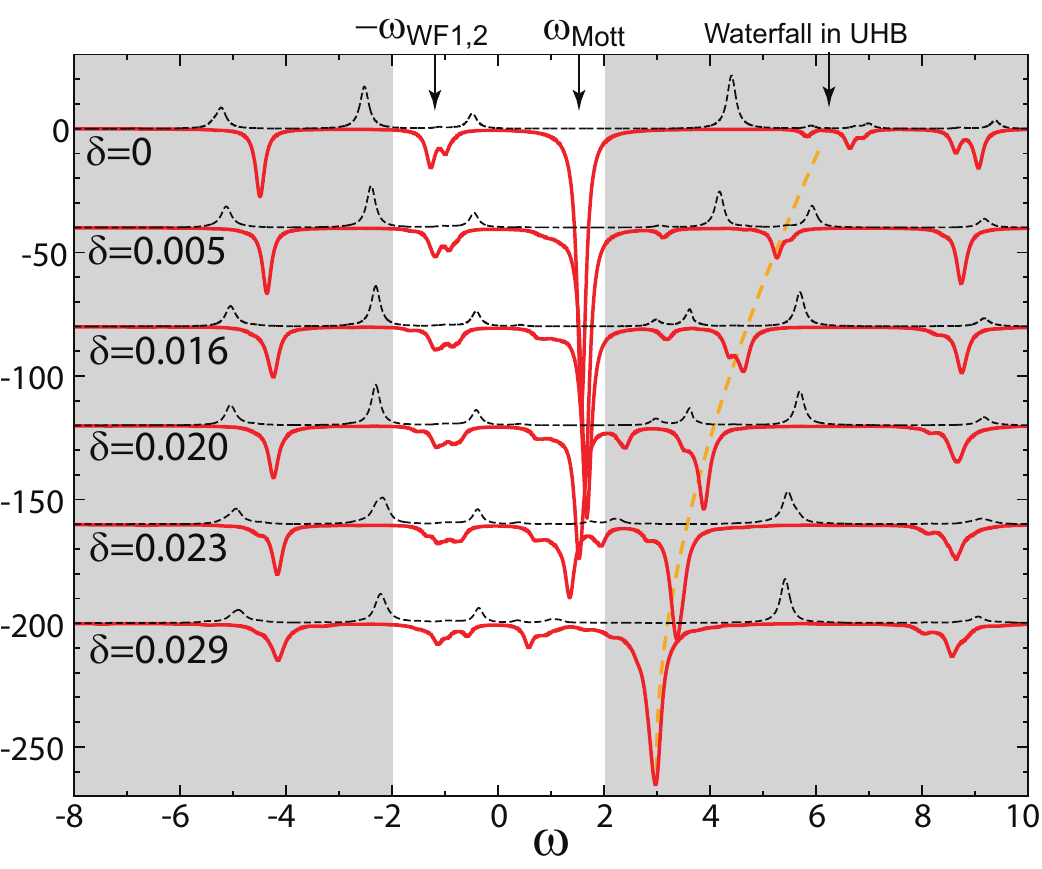}}
\caption{Same as Fig.~\ref{fig:gl7} but for $U=9$. Each curve is shifted by -40 along the vertical axis. }
\label{fig:gl9}
\end{figure}

Figures \ref{fig:gl7}, \ref{fig:gl8} and \ref{fig:gl9} plot Im$\Snor(\kAN,\w)$ and $A(\kAN,\w)$ in a global energy range. In the shaded high-energy area ($|\w|>2$), the doping evolution of these functions is qualitatively similar for $U=7, 8$ and $9$. For $\w<-2$ or $\w\gtrsim U$, there is no significant change with doping. On the other hand, for $2<\w\lesssim U$ we find a notable change with doping. In particular, as indicated by an orange dashed curve, a self-energy peak develops with doping and it acquires a dominant weight at a substantial doping (bottom plots in each figure). This self-energy peak gives the large spectral gap between the ingap state and the UHB.
Interestingly, this self-energy peak traces back to the waterfall structure in the UHB at $\d=0$, as the orange curves indicate. Note that the waterfall structure is present both in the LHB and UHB of the Mott insulator \cite{sakai09PRL,sakai10}, as one can easily understand by considering the electron-hole symmetric case of $t'=0$.

This observation is relevant to the well-known spectral-weight transfer induced by doping  the Mott insulator; the spectral weight is transferred from the UHB to a low energy just above the Fermi level, constituting the ingap state\cite{eskes91}.
Because this ingap state is always located below the self-energy peak pointed out above, it traces back to the weight just below the UHB waterfall at $\d=0$.

In analogy with the waterfall in the LHB (Sec.~\ref{sec:discuss}), the waterfall in the UHB represents an electron addition to the state with dynamically-generated doublon-hole pairs, When the system is doped with holes, such an electron can be added to a hole site, at a significantly lower excitation energy. The resultant ingap state therefore involves the doublon-hole pairs. This is consistent with the interpretation in Sec.~\ref{sec:discuss} because the ingap state is an antibonding state between a low-energy electron and the hidden fermion, which is an electron constituting a doublon weakly bound to a hole.

We note that this reconstruction of the electronic structure in a global energy range is consistent with that obtained previously in the normal-state calculation (see Fig.~1 in Ref.~\onlinecite{sakai09PRL}), too.

\section*{Appendix B: $\text{Im}\Snor$ and $\Akw$ along symmetry lines}

\begin{figure}[tb]
\center{
\includegraphics[width=0.48\textwidth]{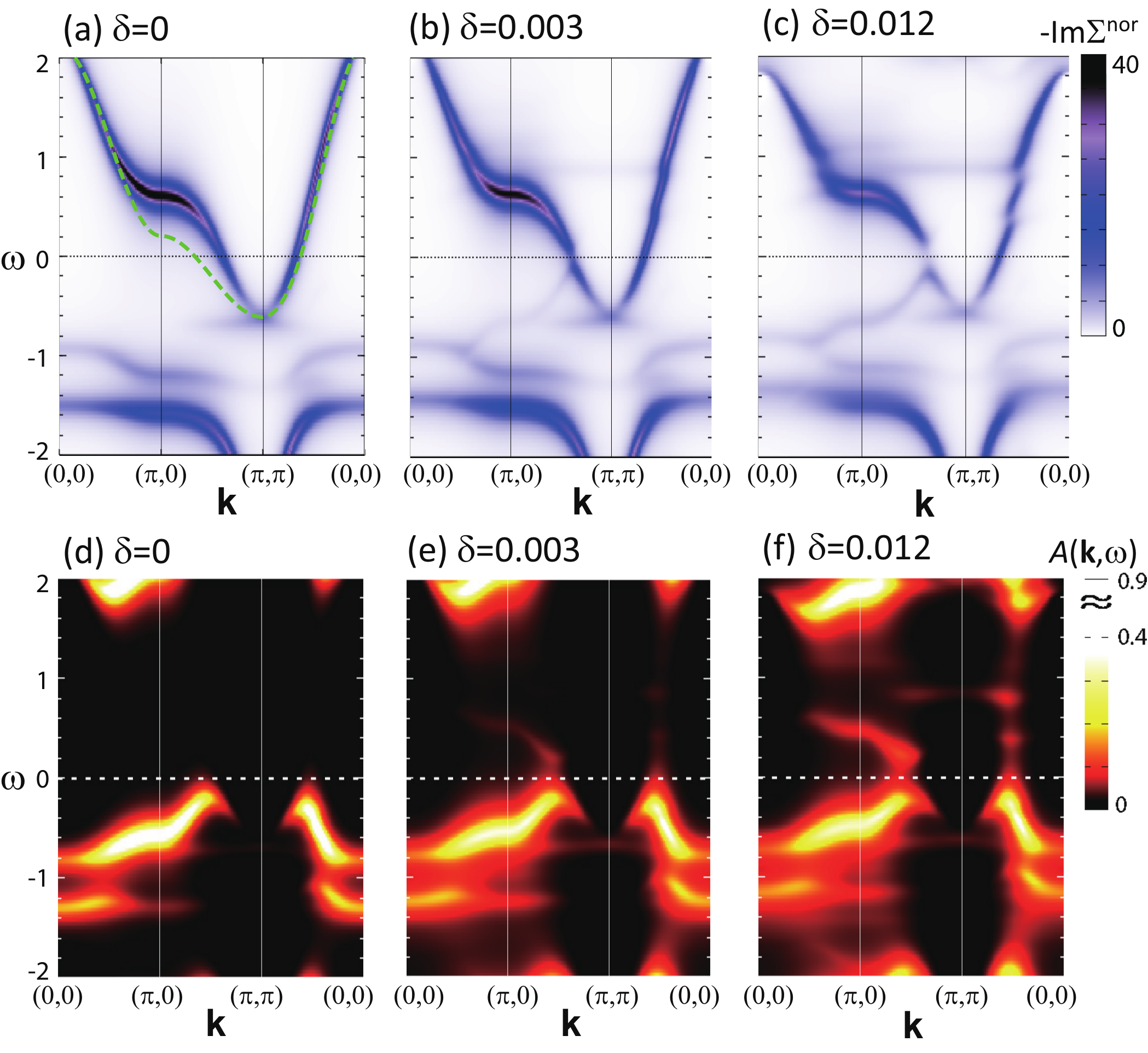}}
\caption{(a),(b),(c) Intensity plot of -Im$\Snor$ along $(0,0)-(\pi,0)-(\pi,\pi)-(0,0)$ for $U=7$. (d),(e),(f) Corresponding plots of $A(\Vec{k},\w)$.
The green dashed curve in panel (a) plots Eq.~(\ref{eq:efmott}) for $\tilde{z}=0.34$ and $\tilde{\mu}=-0.48$.}
\label{fig:skw7}
\end{figure}

\begin{figure}[tb]
\center{
\includegraphics[width=0.48\textwidth]{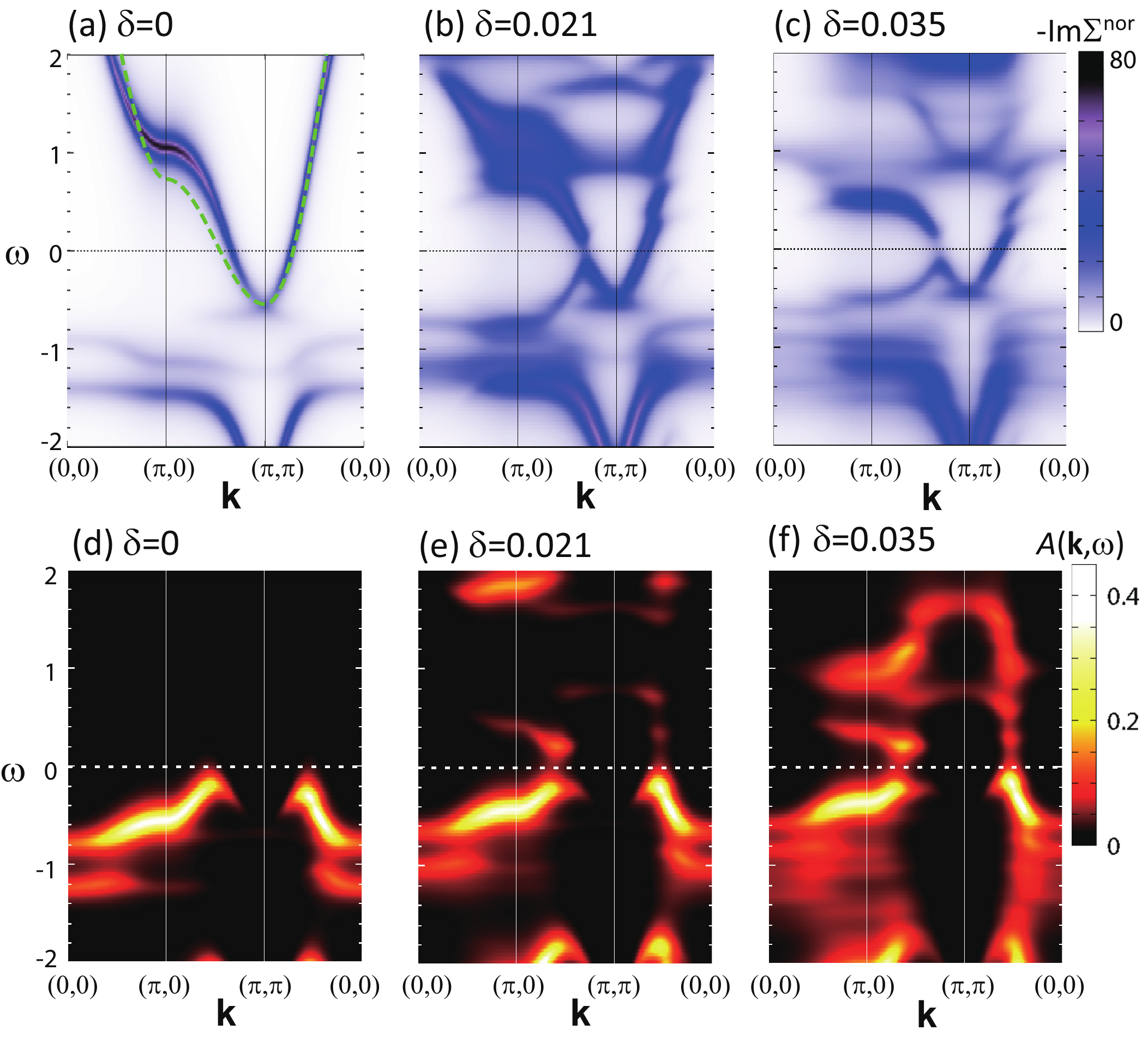}}
\caption{The same as Fig.~\ref{fig:skw7} but for $U=8$. For the green dashed curve in panel (a), we use $\tilde{z}=0.53$ and $\tilde{\mu}=-1.15$.}
\label{fig:skw8}
\end{figure}

\begin{figure}[tb]
\center{
\includegraphics[width=0.48\textwidth]{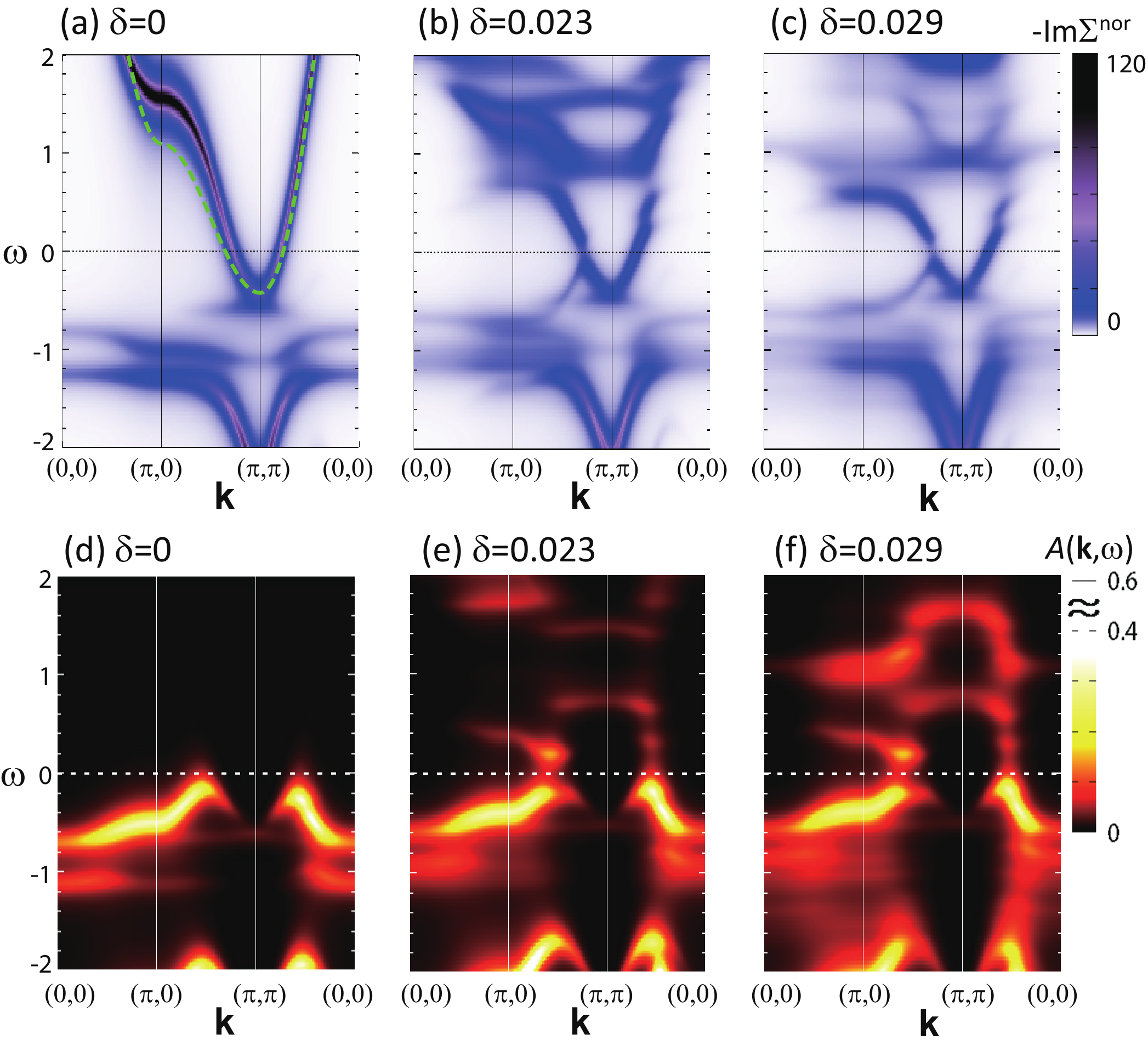}}
\caption{The same as Fig.~\ref{fig:skw7} but for $U=9$. For the green dashed curve in panel (a), we use $\tilde{z}=0.63$ and $\tilde{\mu}=-1.61$.}
\label{fig:skw9}
\end{figure}

Figures \ref{fig:skw7}, \ref{fig:skw8}, and \ref{fig:skw9} show the doping evolution of -Im$\Snor$ and $\Akw$ for $U=7$, 8, and 9, respectively along the $(0,0)-(\pi,0)-(\pi,\pi)-(0,0)$ lines. 
Around $(0,0)$ the self-energy pole generating the Mott gap at $\d=0$ is located at a high energy ($\w>2$) and the weight transfer to the hidden fermion at a tiny doping if any is small and invisible in the figures. The specral weight does not change with doping appreciably in this region.

Around $(\pi,0)$ the Mott-gap peak is located at a lower energy. In fact, for $U=7$, $\wmott$ is smaller than $\wwfa$, and the Mott-gap peak directly transforms into the hidden fermion with doping, keeping its energy position at $\w\simeq0.6$.
On the other hand, for $U=8$ and 9, $\wwfa$ determines the hidden-fermion energy, and a part of the weight at $\w=\wmott$ descends to this energy with doping.
According to these drastic changes of the self-energy, the spectral weight in this region also changes considerably, forming the Boboliubov band and another band just above it. Note that the UHB is located at $\w>2$ for $U=8$ and 9.

Around $(\pi,\pi)$, the Mott-gap peak is below the Fermi level and this low-energy structure does not change appreciably with doping. This makes the spectral weight for $\w<0$ always weak in this region.

Combining these results with Figs.~\ref{fig:sigma7}, \ref{fig:sigma9} and \ref{fig:sigma8}, we conclude that the tiny doping alters the low-energy structure mainly around the $(\pi,0)-(\frac{\pi}{2},\frac{\pi}{2})$ line, where $\wmott$ stays around $\w\sim t$ and the Fermi surface in the normal state appears at a finite doping.

In panel (a) of each figure, we have also plotted a curve defined by 
\begin{align}
\tilde{\e}_{\fmott}(\Vec{k})=\tilde{z}[\e(\Vec{k}+(\pi,\pi))+\mu]-\tilde{\mu},
\label{eq:efmott}
\end{align}
which represents a $(\pi,\pi)$-displaced dispersion of $\fmott$ mentioned in Sec.~\ref{sec:discuss}.
Here $\tilde{z}$ is a renormalization factor, which is taken to be momentum independent for simplicity, and $\tilde{\mu}$ is the onsite energy of $\fmott$.
We determine $\tilde{z}$ and $\tilde{\mu}$ to reproduce the peak positions of Im$\Snor$ at $(0,0)$ and $(\pi,\pi)$ (i.e., the top and bottom of the dispersion).
The dashed green curve indeed reproduces well the overall dispersion of Im$\Snor$ while a discrepancy remains around $(\pi,0)$. 
This discrepancy is attributed to the finite value of $U/t$ since the argument of the $(\pi,\pi)$-displaced dispersion is made in the limit of large $U/t$.

\section*{Appendix C: Results for $U=8$}

\begin{figure}[tb]
\center{
\includegraphics[width=0.48\textwidth]{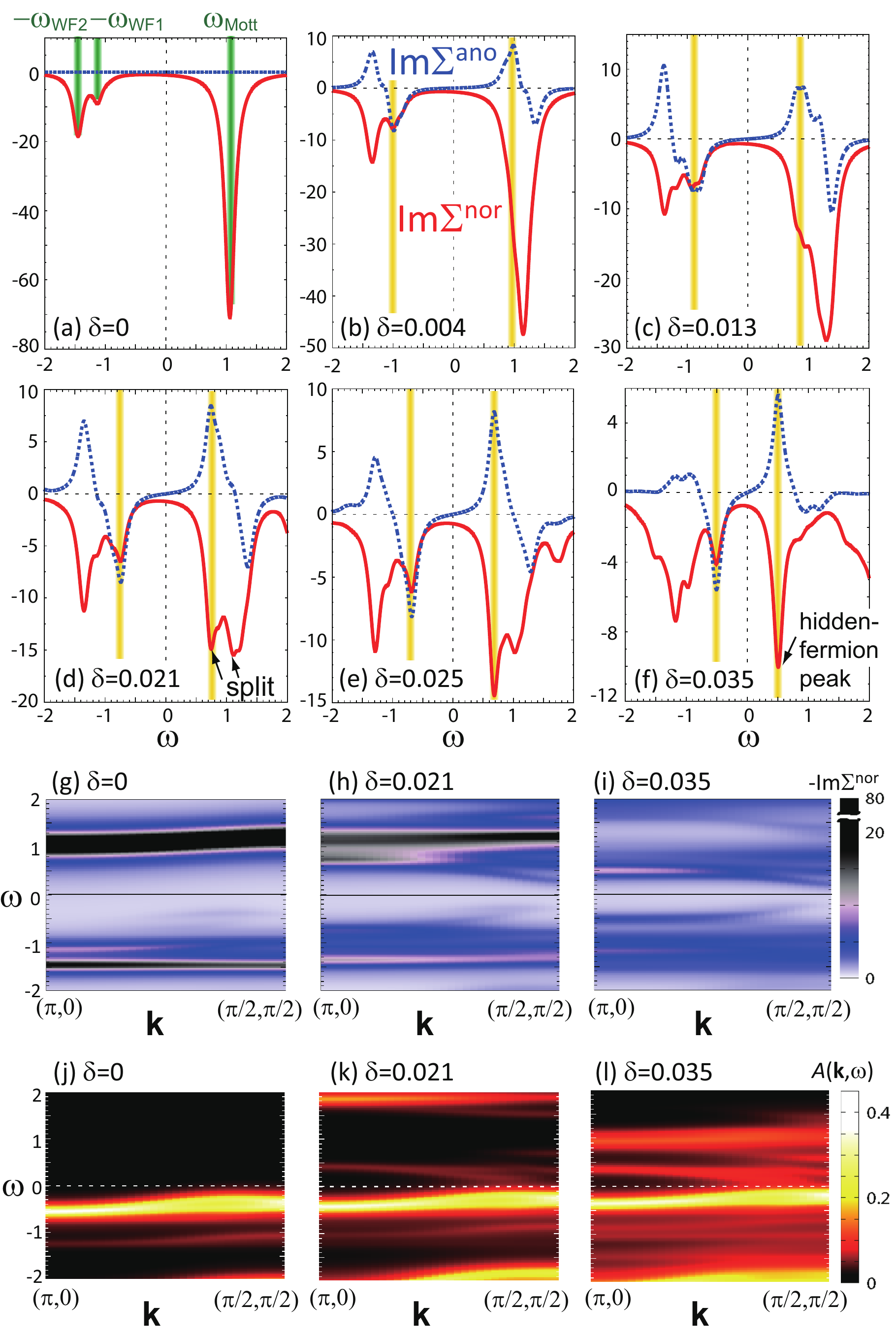}}
\caption{The same as Fig.~\ref{fig:sigma7} but at $U=8$.}
\label{fig:sigma8}
\end{figure}

Figure \ref{fig:sigma8} shows the self-energy and spectral function for $U=8$, where $\wmott$ is comparable to $\w_\text{WF1}$.
In this case, the doping evolution of the self-energy is more involved than the cases for $U=7$ and $U=9$ because of the overlapping of the two energy scales. However, we can still see that the hidden fermion [indicated by a yellow vertical line in Figs.~\ref{fig:sigma8}(b)-(f)] emerges at either $\w=\pm\wwfa$ or $\wmott$.

Because Im$\Sano$ is antisymmetric with respect to $\w$, the doping makes $\Sano$ finite at the same time around $\w=-\w_\text{WF1,2}$ and $\w=\w_\text{WF1,2}$ [Fig.~\ref{fig:sigma8}(b)].
Then, the corresponding structure in Im$\Snor$ at $\w=\w_\text{WF1,2}$ splits the Mott-gap peak into two: The split is evident for $\d>0.013$ [Figs.~\ref{fig:sigma8}(c)(d)(e)]. 
As $\d$ increases, the peak closer to $\w=0$ becomes sharper, with gradually shifting to a lower energy, while the peak at higher frequency loses its weight, which is transferred to an even higher energy.
Eventually in Fig.~\ref{fig:sigma8}(f), the former peak evolves into the hidden-fermion peak of Fig.~\ref{fig:phase}(c).

\section*{Appendix D: Relation between the pole residues of $\Snor$ and $\Sano$}
Suppose that there is only one energetically isolated pole in the low-energy part of the self-energy (This is the case when the hidden-fermion peak has well developed by doping).
Then, the normal and anomalous components of the self-energy is written in the form\cite{sakai16,sakai16PRB},
\begin{align}
\Snor (\w) \simeq& \frac{U}{2}(1-\d) +\frac{V^2(\w+\ef)}{\w^2-\ef^2-D_f^2},\nonumber\\
\Sano (\w) \simeq& D_c+\frac{V^2 D_f}{\w^2-\ef^2-D_f^2}
\label{eq:sig1}
\end{align}
around the pole at $\w=\pm \w_f\equiv \pm\sqrt{\ef^2+D_f^2}$. Here, $D_c$ represents the frequency-independent part of the anomalous self-energy, and $\ef$ and $D_f$ can be interpreted as the energy and the anomalous term of the relevant hidden fermion which hybridizes with electron through $V$. We have abbreviated the momentum argument for the sake of brevity.

The residues of the poles in Eq.~(\ref{eq:sig1}) are easily calculated as
\begin{align}
\text{Res}_\Snor(\w=\pm \w_f)=&\frac{V^2}{2} \left( 1\pm\frac{\ef}{\w_f}\right)\equiv R_{\pm}^\text{nor},\nonumber\\
\text{Res}_\Sano(\w=\pm \w_f)=&\mp\frac{V^2}{2} \frac{D_f}{\w_f} \equiv R_{\pm}^\text{ano}.
\label{eq:res}
\end{align}
Then, we find the following relations between these residues,
\begin{align}
R_{+}^\text{nor}+R_{-}^\text{nor}=&V^2,\nonumber\\
R_{+}^\text{nor}-R_{-}^\text{nor}=&V^2\frac{\ef}{\w_f},\nonumber\\
R_{+}^\text{ano}-R_{-}^\text{ano}=&-V^2\frac{D_f}{\w_f}.
\end{align}
These relations lead to
\begin{align}
(R_{+}^\text{nor}-R_{-}^\text{nor})^2+(R_{+}^\text{ano}-R_{-}^\text{ano})^2=(R_{+}^\text{nor}+R_{-}^\text{nor})^2,
\end{align}
or more simply,
\begin{align}
R_{+}^\text{nor}R_{-}^\text{nor}=(R_{\pm}^\text{ano})^2.
\label{eq:rel}
\end{align}
With the self-energy matrix of Eq.~(\ref{sig}), the above equation can also be written as
\begin{align}
\det \left[\lim_{\w\to\pm \w_f} (\w\mp \w_f)\hat{\Sigma}(\Vec{k},\w)\right]=0.
\label{eq:rel2}
\end{align}
Note that Eq.(\ref{eq:rel}) can also be devived from Eqs.~(20) and (21) in Ref.~\onlinecite{gull15}.
Equation (\ref{eq:rel}) implies that, as far as a total amplitude $R_{+}^\text{nor}+R_{-}^\text{nor}$ is fixed, the product $R_{+}^\text{nor}R_{-}^\text{nor}$ is maximized when $R_{+}^\text{nor}=R_{-}^\text{nor}$ (i.e., electron-hole symmetry) holds.
In fact, around the optimal doping, the self-energy becomes nearly electron-hole symmetric at low energy, as one can see in Figs.~10(b) or 11(a) of Ref.~\onlinecite{sakai16PRB}.

\section*{Appendix E: Fitting of self-energy}

\begin{figure}[tb]
\center{
\includegraphics[width=0.48\textwidth]{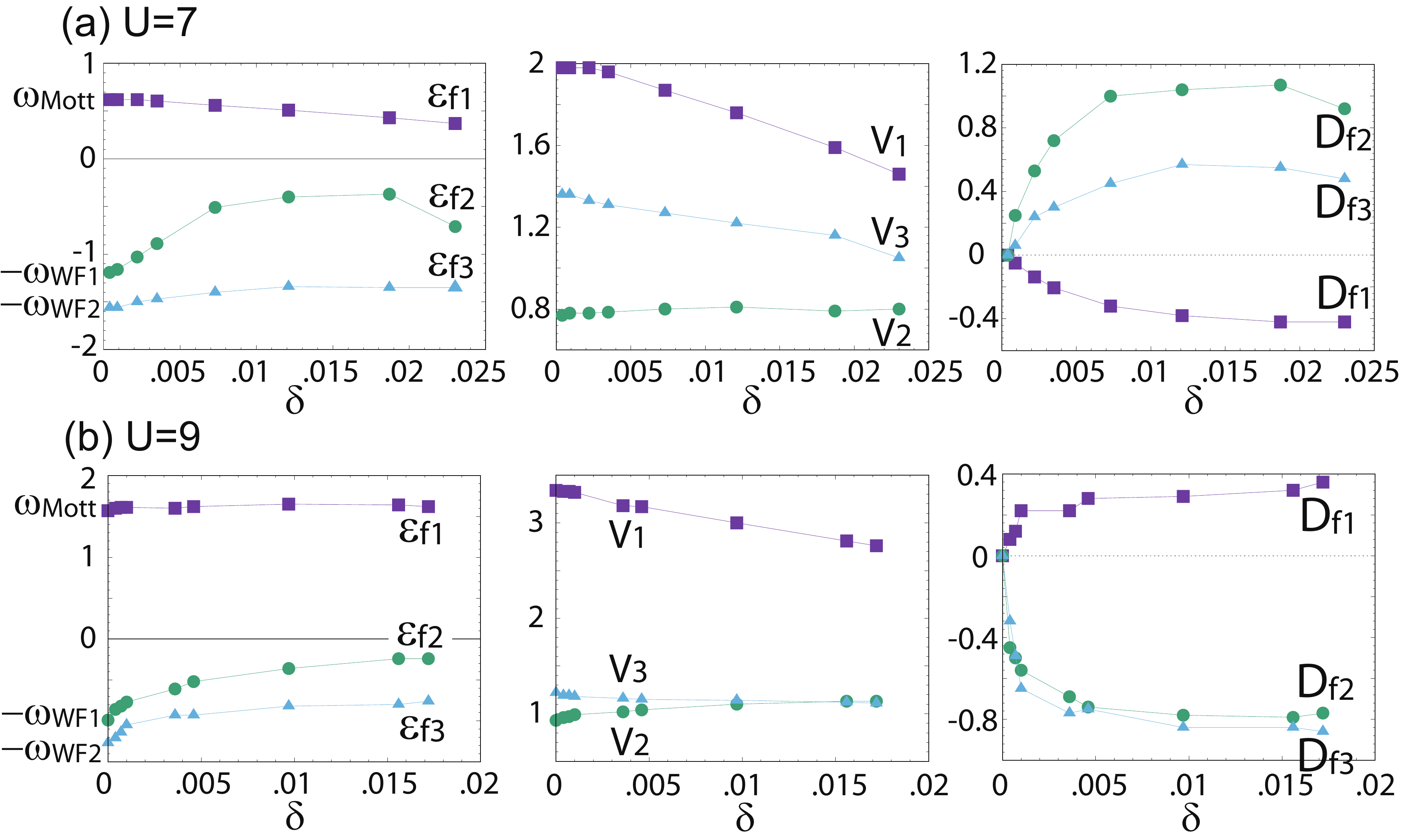}}
\caption{Doping dependence of the fitting parameters. Im$\Snor(\kAN,\w)$ and Im$\Sano(\kAN,\w)$ obtained by the CDMFT are fitted through Eq.~(\ref{eq:sig3}) for (a) $U=7$ and (b) $U=9$.}
\label{fig:para}
\end{figure}

As we have seen in Sec.~\ref{sec:result}, Im$\Snor(\kAN,\w)$ at $\d=0$ shows three peaks at $\w=\wmott$ and $-\w_\text{WF1,2}$. We can then expect that the low-energy part of the self-energy at small $\d$ can be well expressed by the following form,
\begin{align}
\Snor (\w) \simeq& \frac{U}{2}(1-\d) +\sum_{\a=1,3}\frac{V_\a^2(\w+\e_{f_\a})}{\w^2-\e_{f_\a}^2-D_{f_\a}^2},\nonumber\\
\Sano (\w) \simeq& D_c+\sum_{\a=1,3}\frac{V_\a^2 D_{f_\a}}{\w^2-\e_{f_\a}^2-D_{f_\a}^2},
\label{eq:sig3}
\end{align}
which is an extension of Eq.~(\ref{eq:sig1}) \cite{sakai16PRB}. These equations indeed well fit the low-energy part of the self-energy calculated by the CDMFT for $\d\lesssim 0.02$.
At $\d=0$, $\e_{f1}$ agrees with $\wmott$ while $\e_{f2}$ and $\e_{f3}$ agree with $-\wwfa$ and $-\wwfb$, respectively.

Figure \ref{fig:para} shows the obtained fitting parameters for $U=7$ and $U=9$. We find that the $\d$ dependences of $V_\a$ and $D_{f_\a}$ are qualitatively similar for both $U=7$ and $U=9$: As $\d$ increases, $V_1$ decreases while $V_2$ and $V_3$ slightly increases  and decreases, respectively. $|D_{f_\a}|$ rapidly increases at low doping, with keeping $D_{f_1}$ and $D_{f_{2,3}}$ to be different signs. Note that the overall sign of $\{ D_{f_\a} \}$ does not matter because of the $d$ symmetry of the pairing.

A qualitative difference between $U=7$ and $U=9$ cases is in the magnitude relation between $\e_{f1}$ and $\e_{f2,3}$: For $U=7$ $|\e_{f1}|$ is smaller than $|\e_{f2,3}|$ at least for small $\d$ while for $U=9$ $|\e_{f1}|$ is always larger than $|\e_{f2,3}|$.
 This difference produces the different appearances of the self-energy evolution discussed in Secs.~\ref{ssec:u7} and \ref{ssec:u9}.

The sign of $\e_{f_\a}$ is related to the electron-hole asymmetry between $R_{+}^\text{nor}$ and $R_{-}^\text{nor}$, as one can easily see with Eq.~(\ref{eq:res}). 
For $U=7$, the hidden fermion $f_1$ enhancing the superconductivity emerges from $\wmott$ so that $\e_{f_1}$ is positive, leading to $R_{+}^\text{nor} > R_{-}^\text{nor}$. 
On the other hand, for $U=9$ the hidden fermion $f_2$ emerges from $-\wwfa$ so that $\e_{f_2}$ is negative, leading to $R_{+}^\text{nor} < R_{-}^\text{nor}$ at tiny dopings. As Fig.~\ref{fig:para}(b) shows, this negative $\e_{f_2}$ approaches zero as $\d$ increases, and may change sign for $\d>0.02$, as indicated by the relation $R_{+}^\text{nor} > R_{-}^\text{nor}$ seen in Figs.~\ref{fig:sigma9}(e) and (f). In this region, however, the fitting with the three poles does not work well (though the parameters related to $f_2$ still seem to evolve continuously) so that we avoid to conclude.

\bibliography{ref}

\end{document}